\documentclass[twocolumn,aps,prx,floatfix]{revtex4-2}
\pdfoutput=1
\usepackage[T1]{fontenc}
\usepackage{amsmath,amssymb}
\usepackage{graphicx}
\usepackage[leftcaption]{sidecap}
\usepackage{tikz}
\usetikzlibrary{arrows.meta,decorations.pathreplacing,decorations.pathmorphing}
\usepackage[colorlinks,allcolors=blue]{hyperref}

\usepackage[most]{tcolorbox}
\newtcbox{\othermathbox}[1][]{nobeforeafter, math upper, tcbox raise base, 
          enhanced, rounded corners, colback=black!5, colframe=black,
          left=0.7em, top=0.4em, right=0.7em, bottom=0.5em}
\usepackage{empheq}

\newcommand{\be}{\begin{equation}}
\newcommand{\ee}{\end{equation}}
\newcommand{\AAA}{M}
\newcommand{\bA}{\textbf A}
\newcommand{\bB}{\textbf B}
\newcommand{\bp}{\textbf p}
\newcommand{\bbr}{\textbf r}

\newcommand{\cc}{c}
\newcommand{\CC}{\mathbb{C}}
\newcommand{\cH}{{\cal H}}
\newcommand{\cO}{O}

\newcommand{\dd}{\text{d}}
\newcommand{\e}{\text{e}}
\newcommand{\eg}{{\it e.g.}\ }

\DeclareMathOperator{\erfc}{erfc}
\newcommand{\hamm}{H_A}
\newcommand{\ie}{{\it i.e.}\ }

\newcommand{\N}{N}

\newcommand{\nord}[1]{\,:\mathrel{#1}:}
\newcommand{\rhom}{\rho_A}
\newcommand{\Q}{Q}
\newcommand{\RR}{\mathbb{R}}
\newcommand{\ZZ}{\mathbb{Z}}
\newcommand{\bra}[1]{\langle #1|}
\newcommand{\ket}[1]{|#1 \rangle}

\begin{document}

\title{Equipartition of Entanglement in Quantum Hall States}

\author{Blagoje Oblak}
\affiliation{Sorbonne Universit\'e, CNRS, Laboratoire de Physique Th\'eorique et Hautes Energies, LPTHE, F-75005 Paris, France;\\
CPHT, CNRS, Ecole Polytechnique, IP Paris, F-91128 Palaiseau, France}
\author{Nicolas Regnault}
\affiliation{Department of Physics, Princeton University, Princeton, New Jersey 08544, USA}
\author{Benoit Estienne}
\affiliation{Sorbonne Universit\'e, CNRS, Laboratoire de Physique Th\'eorique et Hautes Energies, LPTHE, F-75005 Paris, France}

\begin{abstract}
We study the full counting statistics (FCS) and symmetry-resolved entanglement entropies of integer and fractional quantum Hall states. For the filled lowest Landau level of spin-polarized electrons on an infinite cylinder, we compute exactly the charged moments associated with a cut orthogonal to the cylinder's axis. This yields the behavior of FCS and entropies in the limit of large perimeters: in a suitable range of fluctuations, FCS is Gaussian and entanglement spreads evenly among different charge sectors. Subleading charge-dependent corrections to equipartition are also derived. We then extend the analysis to Laughlin wavefunctions, where entanglement spectroscopy is carried out assuming the Li-Haldane conjecture. The results confirm equipartition up to small charge-dependent terms, and are then matched with numerical computations based on exact matrix product states.
\end{abstract}

\maketitle

\section{Introduction}
\label{sec:INTRO}

Entanglement plays a central role in the construction of many-body quantum states in terms of tensor networks, which provide efficient numerical methods for the simulation of strongly correlated systems \cite{PhysRevLett.69.2863,RevModPhys.77.259,doi:10.1080/14789940801912366,Orus_2014}. It is also a powerful theoretical tool probing quantum correlations, both in condensed matter and in high-energy physics \cite{RevModPhys.80.517,LAFLORENCIE20161}. For instance, the entanglement entropy (EE) of gapped phases of matter obeys an area law reminiscent of black hole entropy \cite{Srednicki_1993,RevModPhys.82.277}, while that of one-dimensional (1D) critical systems exhibits an anomalous logarithmic growth sensitive to the central charge \cite{HOLZHEY1994443,PhysRevLett.90.227902,CC04}. In the quantum Hall effect (QHE) to be studied here, entanglement detects intrinsic topological order \cite{PhysRevLett.96.110404,PhysRevLett.96.110405} and identifies gapless edge modes at the boundary \cite{Estienne_2020,Estienne_2021} or at the interface between different fractional quantum Hall states \cite{Crepel_2019a,Crepel_2019b,PhysRevLett.123.126804}. It is thus a crucial theoretical probe of the properties of topological phases of matter.

The goal of this paper is to compute {\it symmetry-resolved} measures of entanglement in the QHE. Indeed, fluctuations of the local charge associated with some internal symmetry have long been investigated in relation to entanglement \cite{PhysRevB.82.012405,PhysRevB.85.035409,2014JSMTE..10..005P}. The corresponding random variable, known as full counting statistics (FCS) \cite{1993ZhPmR..58..225L,1996JMP....37.4845L}, will feature prominently throughout this work. It captures \eg the Luttinger parameter of 1D systems \cite{PhysRevLett.108.116401,Estienne_2020,Hackenbroich_2021}, detects and counts massless Dirac fermions in 2D \cite{Crepel_2021}, and measures the long-wavelength limit of the structure factor of gapped 2D liquids \cite{estienne2021cornering}. It was recently realized that internal symmetries also provide a natural decomposition of entanglement measures in sectors with definite values of the corresponding charge \cite{Laflorencie_2014,PhysRevLett.91.097902,PhysRevLett.120.200602,2021JHEP...10..067C}. Such symmetry-resolved EEs and their relation to charge fluctuations have been studied in a number of contexts: critical \cite{PhysRevLett.120.200602,PhysRevB.98.041106,PhysRevB.100.235146,PhysRevA.98.032302,Murciano_2021,capizzi2020symmetry,Bonsignori_2020,10.21468/SciPostPhys.10.3.054,Chen_2021,capizzi2021symmetry,PhysRevB.103.L041104} or gapped \cite{PhysRevA.100.022324,Calabrese_2020} 1D systems, topological phases \cite{PhysRevB.99.115429,PhysRevResearch.2.043191,azses2020symmetry}, systems of free particles \cite{PhysRevLett.121.150501,tan2019particle,Bonsignori_2019,Fraenkel_2020,murciano2020entanglement,2020JSMTE2020h3102M,10.21468/SciPostPhys.8.6.083,Horvath_2021,parez2021exact,fraenkel2021entanglement,Crepel_2021}, integrable models \cite{10.21468/SciPostPhys.8.3.046,horvth2020symmetry,horvath2021branch}, and even gravity \cite{Zhao_2021,weisenberger2021symmetryresolved}. As it turns out, entropy typically spreads evenly among different symmetry sectors---a property dubbed {\it equipartition of EE} \cite{PhysRevB.98.041106}.

In this work we investigate the equipartition of entropy and its charge-dependent corrections for integer and fractional quantum Hall states. The latter live on an infinite cylinder whose perimeter is large compared to the magnetic length, and the entangling region is a `half cylinder' whose boundary is perpendicular to the cylinder's axis. Symmetry-resolved EE then satisfies an area law at leading order, while its dependence on charge is captured by subleading terms that vanish in the thermodynamic limit. A notion of universality emerges: charge-dependent corrections obey similar scaling laws in both integer and (Abelian) fractional phases, regardless of the filling fraction. Furthermore, FCS has a leading Gaussian form in both cases. This similarity ultimately stems from the fact that the leading contribution to entanglement originates from edge modes at the boundary of the entangling region, through the bulk-edge correspondence. The only difference is that non-interacting setups allow for a microscopic derivation of all coefficients thanks to standard free fermion methods \cite{Peschel_2009,2009PhRvB..80o3303R,2010JSMTE..12..033R,2012PhRvB..85k5321D}, whereas the fractional QHE requires field-theoretic arguments such as the Li-Haldane conjecture \cite{PhysRevLett.96.110404,Li_2008} and its irrelevant corrections \cite{2012PhRvB..86x5310D}. The resulting predictions for FCS and symmetry-resolved EE are eventually shown to be consistent with exact numerical results obtained using matrix product states (MPSs).

The paper is organized as follows. We start in Sec.\ \ref{seSymmRes} with a brief review of symmetry-resolved entanglement for generic (interacting or free) many-body systems. Sec.\ \ref{sePREL} is then devoted to entanglement spectroscopy in a filled lowest Landau level (LLL) on a cylinder. This is used in Sec.\ \ref{seSRE} to deduce charged moments, FCS and symmetry-resolved EE in the integer QHE. Finally, Sec.\ \ref{seFQHE} focuses on analytical and numerical computations of entanglement in fractional quantum Hall states. We briefly conclude in Sec.\ \ref{sec:CON} with a summary and a list of potential follow-ups, while the \hyperref[app:asymptotics]{Appendices} contain some further computational and numerical details. 

\section{Symmetry-resolved \\ reduced density matrix}
\label{seSymmRes}

This section sets up our notation and serves as a brief reminder of the definition of FCS, charged moments and symmetry-resolved entropies. Consider therefore a many-body quantum system enjoying a global U(1) symmetry whose conserved charge $\Q$ is the integral of some local density. (A typical example below will be the number operator.) Splitting the spatial sample in two subregions $A$ and $B$ yields a bipartition $\cH=\cH_A\otimes\cH_B$ of the system's Hilbert space and  the total charge can be decomposed as $\Q=\Q_A\otimes\mathbb{I}_B+\mathbb{I}_A \otimes \Q_B$  where $\Q_A$ ($\Q_B$) is the charge in region $A$ ($B$). As long as the total density matrix $\rho$ describes a statistical superposition of states with definite charge, it commutes with $\Q$. A partial trace over $\cH_B$ then gives $[\rhom,\Q_A]=0$ where $\rhom\equiv\text{Tr}_B (\rho)$ is the reduced density matrix (RDM) of region $A$. The RDM is thus block-diagonal with respect to $\Q_A$:
\be
\label{e3}
\rhom
=
\bigoplus_q\,
\Pi_q\,\rhom
=
\bigoplus_q\,
p_q\,\rhom(q)\,,
\ee
where $\Pi_q$ is the projector on the eigenspace of $\Q_A$ with eigenvalue $q$, while $p_q\equiv\text{Tr}  (\Pi_q \rhom)$ is the probability that the charge in region $A$ be $\Q_A = q$. This distribution $p_q$ is known as {\it full counting statistics} (FCS) \cite{1993ZhPmR..58..225L,1996JMP....37.4845L}, with characteristic function
\be
\label{e4}
\widehat{Z}_1(\alpha)
\equiv
\big<\e^{i \alpha \Q_A}\big>
\equiv
\text{Tr}\big(\e^{i \alpha \Q_A} \rhom\big)\,.
\ee 
We assume throughout that the admissible values of the charge $Q_A$ are integers up to some real constant shift $\delta$, \ie $q\in\ZZ+\delta$, as will be the case for charge deviations in the QHE.

The resolution \eqref{e3} of the RDM in symmetry sectors provides a natural refinement of entanglement measures. Each block $\rhom(q)$ is indeed normalized as $\text{Tr}\rhom(q)=1$: it is a {\it bona fide} RDM that actually coincides with the collapsed RDM following a measurement $\Q_A = q$. The {\it symmetry-resolved} von Neumann entanglement entropy (EE) at charge $q$ is thus defined as
\be
\label{e5}
S_1(q)
\equiv
-\text{Tr}\big(\rhom(q)\log\rhom(q)\big)\,,
\ee
encoding the amount of entanglement in sector $q$. A straightforward computation then shows that the total EE, $S_1\equiv-\text{Tr}(\rhom\log\rhom)$, can be split in two terms:
\be
\label{vndec}
S_1
=
-\sum_q p_q\log p_q
+\sum_q p_q\,S_1(q)\,. 
\ee
The first contribution is simply the Shannon entropy of charge fluctuations, while the second is the average entropy per sector. This decomposition strongly suggests the following thought experiment for measuring the quantum state of $A$: one can first measure the charge $\Q_A$ with some outcome $q$, collapsing the RDM to $\rhom(q)$, and only then measure the state of subsystem $A$. Other entanglement measures can also be refined, such as symmetry-resolved R\'enyi entropies
\be
\label{e7}
S_n(q)
\equiv
\frac{1}{1-n}
\log\text{Tr}\big(\rhom(q)^n\big)\,,
\qquad
n>1.
\ee
On general grounds, the total EE is expected to be distributed evenly among different charge sectors \cite{Laflorencie_2014}. Such an `equipartition of EE' has been shown to hold in many 1D systems \cite{PhysRevB.98.041106,10.21468/SciPostPhys.8.3.046,turkeshi2020entanglement}, as well as some free 2D setups \cite{2020JSMTE2020h3102M}. To be precise, equipartition is only true for $q-\langle \Q_A\rangle$ much smaller than the standard deviation of $\Q_A$; for large fluctuations, symmetry-resolved entropy becomes arbitrarily small instead. We will similarly encounter equipartition of entanglement in the QHE, with suitable charge-dependent corrections.

The computation of symmetry-resolved entropies is facilitated by introducing generating functions that generalize the characteristic function \eqref{e4} of FCS:
\be
\label{zefour}
\widehat{Z}_n(\alpha)
\equiv
\text{Tr}\big(\e^{i \alpha \Q_A} \rhom ^n\big)\,.
\ee
These quantities are known as {\it charged moments} \cite{PhysRevLett.120.200602}, or `charged R\'enyi entropies' in a different context \cite{Belin_2013,Pastras_2014,Belin_2015,PhysRevB.93.195113,PhysRevD.93.105032,Dowker_2016,PhysRevD.96.065016,Dowker_2017}. They satisfy $\widehat{Z}_n(\alpha+2\pi)=\e^{2\pi i\delta}\widehat{Z}_n(\alpha)$ when the spectrum of $\Q_A$ lies in $\ZZ+\delta$, so we restrict attention to $\alpha\in\,]-\pi,\pi[$ from now on. All symmetry-resolved entropies can be extracted from such charged moments. Indeed, notice first that their Fourier transform is 
\be
\label{zen}
Z_n(q)
\equiv
\int_{-\pi}^{\pi}
\frac{\dd\alpha}{2\pi}
\e^{-i \alpha q} \widehat{Z}_n(\alpha)
= 
\text{Tr}\big(\Pi_q\,\rhom ^n\big)
\ee
where $\Pi_q$ is the projector of \eqref{e3}, as follows from the operator identity $\Pi_q
=\int_{-\pi}^{\pi}\dd\alpha\,\e^{i \alpha (\Q_A-q)}/(2\pi)$. The FCS defined below \eqref{e3} and the symmetry-resolved R\'enyi entropies \eqref{e7} are then recovered as 
\be
\label{fcs}
p_q
=
Z_1(q),
\qquad
S_n(q)
=
\frac{1}{1-n}\log\frac{Z_n(q)}{Z_1(q)^n}\,,
\ee
while the von Neumann entropy \eqref{e5} in sector $q$ reads
\be
\label{e12}
S_1(q)
=
- \frac{\dd}{\dd n}
\frac{Z_n(q)}{Z_1(q)^n}
\bigg|_{n=1}\,.
\ee
In what follows we compute these objects in integer and fractional QHEs (Secs.~\ref{sePREL}--\ref{seSRE} and \ref{seFQHE}, respectively).

\section{Entanglement of free fermions on a cylinder}
\label{sePREL}

Here we recall how the entanglement spectrum of the $\nu=1$ quantum Hall state can be found using  standard free fermion methods \cite{Peschel_2009,2009PhRvB..80o3303R,2010JSMTE..12..033R,2012PhRvB..85k5321D}. We then write down exact formulas for charged moments \eqref{zefour}. This is a key prerequisite for Sec.\ \ref{seSRE} on FCS and symmetry-resolved EE. 

\subsection{Entanglement spectrum of a filled LLL}
\label{seLLL}

\begin{figure}[t]
\centering
\includegraphics[width=0.95\columnwidth]{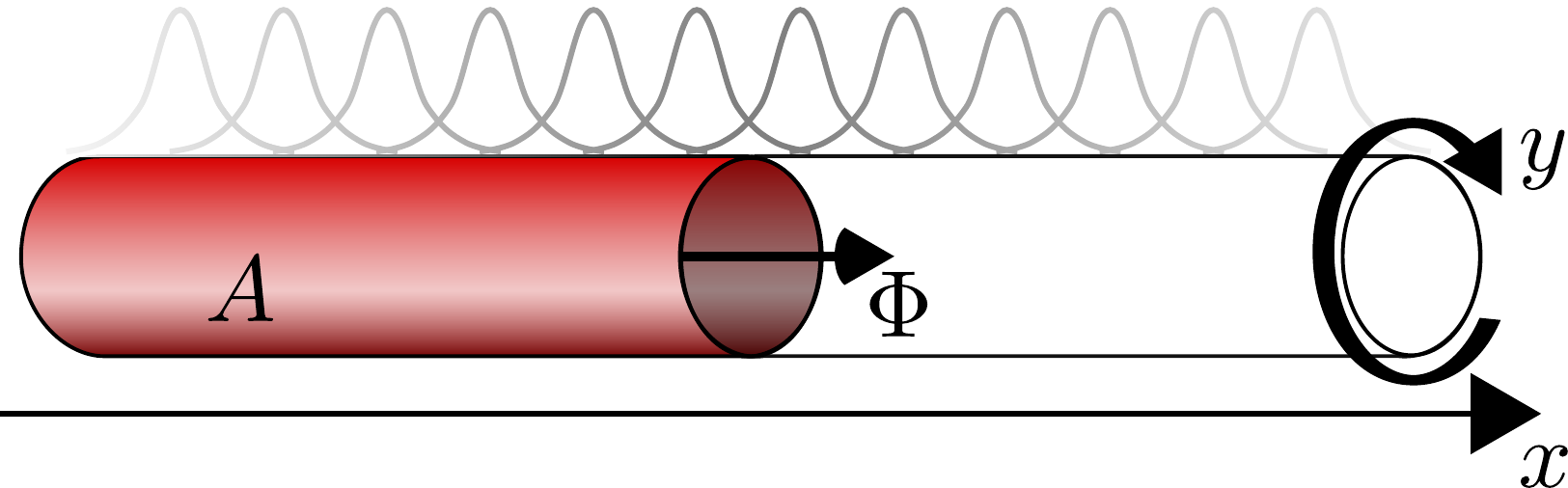}
\caption{The geometric setup of this paper. Electrons live on an infinite cylinder (with coordinates $x,y$) whose subregion $A$ consists of points with $x<0$. The dimensionless flux $\Phi$ through the boundary of $A$ affects entanglement, since it changes the location of maxima of LLL wavefunctions \eqref{phim}.}
\label{ficyl}
\end{figure}

Consider non-interacting spinless electrons confined to a cylinder $\RR\times S^1$ whose points $\bbr$ are labeled by coordinates $(x,y)$, where $y\sim y+L$ in terms of a perimeter $L$ (see Fig.\ \ref{ficyl}). The cylinder is flat (with line element $\dd s^2 = \dd x^2 + \dd y^2$) and supports a uniform magnetic field $\bB=\dd x\wedge\dd y$, where units were chosen so that cyclotron frequency and magnetic length are set to unity. Each electron is governed by the Landau Hamiltonian $ H=(\bp-\bA)^2/2$ where $\bA=x\,\dd y$ is the vector potential in Landau gauge. The presence of an Aharonov-Bohm flux $\Phi$ piercing through the cylinder implies that eigenstates of the translation operator along $y$ take the form $\phi_k(x,y) =\e^{iky}f(x)$ with some momentum $k\in2\pi(\ZZ+\Phi)/L$. In particular, an orthonormal basis of the lowest Landau level (LLL) is given by
\be
\label{phim}
\phi_k(x,y)
=
\tfrac{1}{\sqrt{L\sqrt{\pi}}}
\e^{iky}\e^{-(x-k)^2/2},
\quad
k\in\tfrac{2\pi}{L}(\ZZ+\Phi)\,.
\ee
The ground state at filling fraction $\nu =1$ then is the Slater determinant of the LLL: $\ket{\Omega}=\bigwedge_{k\in\frac{2\pi}{L}(\ZZ+\Phi)}\ket{\phi_k}.$ 

The full entanglement spectrum of the integer QHE can be computed exactly \cite{2009PhRvB..80o3303R,2012PhRvB..85k5321D} when the system's bipartition enjoys a spatial symmetry (\eg rotations or translations). Accordingly, we decompose the total (many-body) Hilbert space as $\cH=\cH_A\otimes\cH_B$ by splitting the cylinder $\RR\times S^1$ in two subregions $A$ and $B$, where $A=\,]-\infty,0[\,\times S^1$ is the left `half-cylinder' $x < 0$ and $B$ is its complement (see Fig.\ \ref{ficyl}). This choice of bipartition preserves the translational symmetry along $y$, thus allowing an analytical derivation of the entanglement spectrum. For more generic bipartitions, exact computations are out of reach and one can, at best, compute the asymptotics of the entanglement spectrum in the thermodynamic limit using semi-classical methods \cite{2019CMaPh.376..521C}.

For non-interacting fermionic systems such as the integer QHE studied here, powerful techniques are available to evaluate both FCS \cite{Abanov_2011} and (symmetry-resolved) EEs \cite{Jin-Korepin,Bonsignori_2019,Fraenkel_2020}. Indeed, the RDM is a Gaussian state \cite{2003JPhA...36L.205P}
\be
\label{e18}
\rhom
\equiv
\frac{1}{Z} \e^{-\hamm}\,,
\qquad
Z
\equiv
\text{Tr}\,\e^{-\hamm}
\ee
whose modular Hamiltonian $\hamm$ is {\it quadratic} in terms of local fermionic creation and annihilation operators in $A$, say $\Psi^{\dagger}(\bbr)$ and $\Psi(\bbr)$ \cite{PhysRevB.64.064412,2003JPhA...36L.205P,Peschel_2009,2012BrJPh..42..267P}. The problem of computing the RDM thus reduces to a one-body exercise: diagonalizing the  kernel of $\hamm$. Since the latter is unknown {\it a priori}, one more trick is required: starting from the correlation function \footnote{The correlation function \eqref{e20} is nothing but the Bergman kernel in the $\nu=1$ QHE \cite{2010CMaPh.293..205D,2019CMaPh.376..521C}.}
\be
\label{e20}
\CC(\bbr,\bbr')
\equiv
\langle\Omega|\Psi^{\dagger}(\bbr')\Psi(\bbr)|\Omega\rangle
=
\sum_{k}
\langle\bbr\ket{\phi_k}\bra{\phi_k}\bbr'\rangle\,,
\ee
one can reverse-engineer Wick's theorem to reconstruct $\hamm$ from the knowledge of correlations in $A$ \cite{2003JPhA...36L.205P,PhysRevB.64.064412}. To do so, consider the self-adjoint operator
\be
\label{e21}
 C_A
\equiv
\iint\limits_{A\times A}
\dd^2\bbr\,\dd^2\bbr'\,
\CC(\bbr,\bbr')
|\bbr\rangle\langle\bbr'|
=
 P_A  P_{\Omega}  P_A
\ee
acting on one-body states localized in $A$, where $ P_A\equiv\int_A\dd^2\bbr\ket{\bbr}\bra{\bbr}$ is the projector on region $A$ and $ P_{\Omega}\equiv\sum_{k}\ket{\phi_k}\bra{\phi_k}$ is the projector on occupied states (here the LLL). $C_A$ then has some single-particle eigenstates $\ket{\psi_m}$ with eigenvalues $\lambda_m$ say, which turn out to coincide with eigenstates of the (kernel of the) modular Hamiltonian. Indeed, introducing fermionic annihilation operators $\cc_m^{\dagger}\equiv\int_A\dd^2\bbr\,\psi_m(\bbr)\Psi^{\dagger}(\bbr)$ for $\ket{\psi_m}$'s, the tautological identity $\CC(\bbr,\bbr')=\text{Tr}(\rhom\,\Psi^{\dagger}(\bbr')\Psi(\bbr))$ for $\bbr,\bbr'\in A$ leads to $\text{Tr}(\rhom\cc_m^{\dagger}\cc_n)=\bra{\psi_n} C_A\ket{\psi_m}=\delta_{mn}\lambda_m$. The rewriting \eqref{e18} of the RDM as the exponential of a modular Hamiltonian then yields the diagonal expression
\be
\label{hamm_diagonal}
\hamm
=
\sum_m
\epsilon_m\,
\cc_m^{\dagger} \cc_m\,,
\qquad
\epsilon_m
\equiv
\log\Big(\frac{1-\lambda_m}{\lambda_m}\Big),
\ee
with eigenvalues $\epsilon_m$ that we refer to as {\it pseudo-energies}. Note that the eigenvalues $\lambda_m$ of $ C_A= P_A P_{\Omega} P_A$ belong to the interval $[0,1]$, with boundary values $0$ and $1$ that are irrelevant: they correspond to infinite pseudo-energies that do not contribute to the entanglement spectrum.

If one is interested in the spectrum without regard to eigenstates, an alternative approach is to consider the overlap matrix of occupied states \eqref{phim} \cite{2006JPhA...39L..85K,2009PhRvB..80o3303R,2011PhRvL.107b0601C}, 
\be
\label{e28}
\AAA_{mn}
\equiv
\int_A\dd^2\bbr\,
\phi_{k_m}^*(\bbr)\phi_{k_n}(\bbr)\,,
\quad k_m
\equiv
\frac{2\pi}{L}(m+\Phi)
\ee
with $m\in\ZZ$. The entries of \eqref{e28} are matrix elements of the operator $ O_A\equiv P_{\Omega} P_A P_{\Omega}$. The point is that all non-zero eigenvalues of $ O_A$ and $ C_A$ coincide: they are the squares of singular values of $ P_A P_{\Omega}$, as follows from $ C_A= P_A P_{\Omega}( P_A P_{\Omega})^{\dag}$ and $ O_A=( P_A P_{\Omega})^{\dag} P_A P_{\Omega}$. In our case, dealing with overlaps rather than the correlation operator \eqref{e21} turns out to be simpler, as this trades the diagonalization of a continuous kernel for that of a discrete matrix. Translation invariance along $y$ actually trivializes the problem since the overlap matrix \eqref{e28} of LLL states \eqref{phim} is automatically diagonal, with eigenvalues $\lambda_m$ labeled by momentum:
\be
\label{ovlap}
\lambda_m
=
\lambda(k_m)\,,
\qquad
\lambda(k)
\equiv
\frac{1}{2}\erfc(k)
\ee
where $\erfc$ is the complementary error function. Note the spectral flow (or charge pumping) that can be read off from \eqref{ovlap}: the spectrum returns to itself when the flux $\Phi$ increases by one unit, mapping $\lambda_m\mapsto\lambda_{m+1}$. 

\begin{figure}[t]
\centering
\includegraphics[width=0.70\columnwidth]{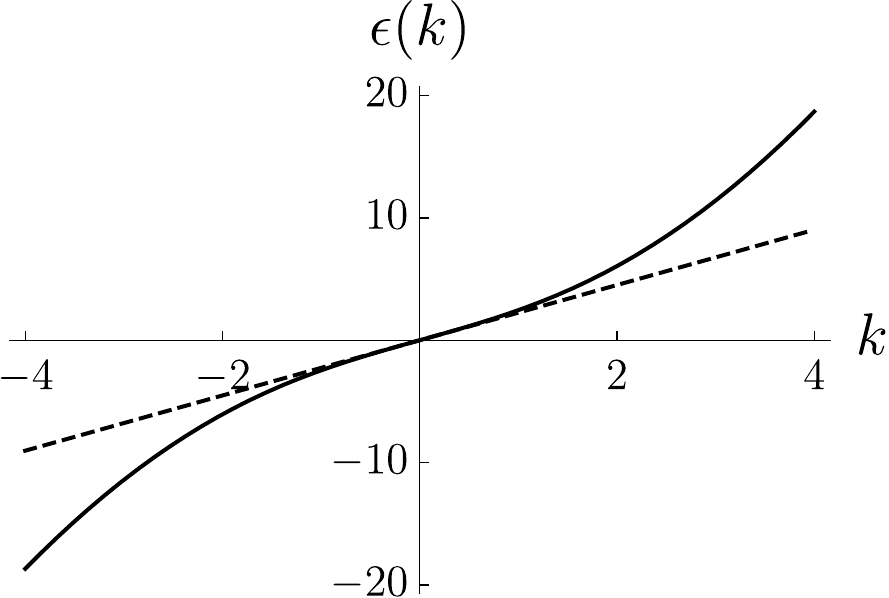}
\caption{The pseudo-dispersion relation \eqref{em} (full line) compared to its low-momentum approximation $\epsilon(k)\sim 4k/\sqrt{\pi}$ (dashed line). This linearization will be used in Sec.~\ref{sec:corrections}.}
\label{fidispersion}
\end{figure}

The eigenvalues $\lambda_m \in\,]0,1[$ control the distribution of the number of particles in region $A$, \ie FCS. The latter is indeed a sum of (infinitely many) independent Bernoulli random variables with parameters $\lambda_m$, $m\in\ZZ$ \cite{2019CMaPh.376..521C}. The overlaps \eqref{ovlap} also give access to the many-body entanglement spectrum, since \eqref{hamm_diagonal} relates pseudo-energies to overlap eigenvalues:
\be
\label{em}
\epsilon_m
=
\epsilon(k_m),
\qquad
\epsilon(k) 
\equiv
\log \frac{\erfc\big({-}k\big)}{\erfc\big(k\big)}\,.
\ee
This dispersion relation is plotted in Fig. \ref{fidispersion}, along with the low-momentum linear approximation $\epsilon(k)\sim4k/\sqrt{\pi}$ that will briefly be used at the end of Sec.~\ref{sec:corrections}. 

\begin{figure}[t]
\centering
\includegraphics[width=0.97\columnwidth]{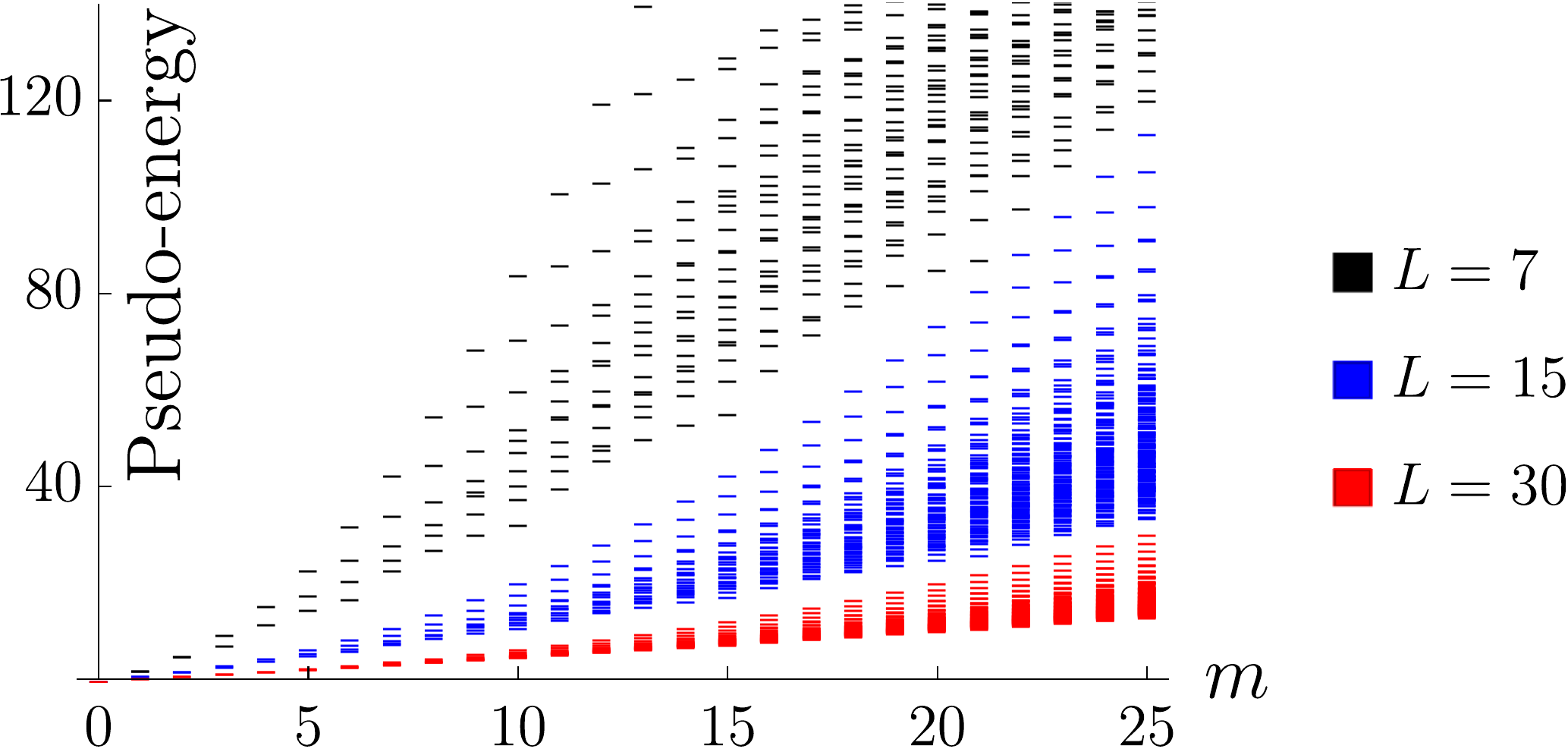}
\caption{Many-body entanglement spectra of a filled LLL on half-cylinders at zero flux with perimeters $L=7,15,30$. Pseudo-energies are normalized with respect to $m=0$ and obtained by adding up one-body energies \eqref{em} in accordance with fermionic statistics. The spectrum flattens and the CFT approximation improves as $L$ increases. Note the similarity with \cite[Figs.\ 1--2]{2012PhRvB..85k5321D} and \cite[Figs.\ 7--8]{2012PhRvB..86x5310D}.}
\label{fispec}
\end{figure}

At this stage a technical subtlety must be addressed: the ground state energy of the modular Hamiltonian \eqref{hamm_diagonal} diverges, as does the normalization $Z=\text{Tr}\,\e^{-\hamm}$, due to the presence of infinitely many modes with negative pseudo-energy. This is easily cured with normal ordering: one may choose to write $\hamm$ as a sum of terms of the form $\cc_m^{\dagger}\cc_m$, but some of these may be recast as $-\cc_m\cc_m^{\dagger}$ up to constants that can be absorbed in $Z$. Accordingly, let
\be
\label{e32}
\hamm
\equiv
\sum_{m\in\ZZ} \epsilon_m : \cc_m^{\dagger} \cc_m :\,,
\ee
where normal ordering is defined by \footnote{Instead of \eqref{eq:nord}, one may define a normal-ordering prescription based on whether $m\geq m_0$ or $m<m_0$ for some integer $m_0\neq0$. The choice is ultimately irrelevant, as changing $m_0$ shifts the modular Hamiltonian by a {\it finite} constant that may be absorbed in the normalization $Z$.}
\be
\label{eq:nord}
\nord{\,\cc_m^{\dagger}\cc_m}\,
\equiv
\begin{cases}
\cc_m^{\dagger}\cc_m & \text{if }m\geq0\,,\\
-\cc_m\cc_m^{\dagger} & \text{if }m<0\,.
\end{cases}
\ee
The normalization constant of \eqref{e18} thus reads
\be
\label{finiz}
Z
\equiv
\text{Tr}\left[\e^{-\sum\limits_{m\in\ZZ}\epsilon_m\nord{\cc_m^{\dagger}\cc_m}}\right]
=
\prod_{m<0}\frac{1}{\lambda_m}\prod_{m\geq0}\frac{1}{1-\lambda_m}
\ee
and is finite as desired since overlaps \eqref{ovlap} converge exponentially to $0$ (resp.\ $1$) for $m\to+\infty$ (resp.\ $m\to-\infty$). The many-body entanglement spectrum then follows from fermionic statistics. It is plotted in Fig.\ \ref{fispec} as a function of momentum, for several values of perimeter $L$. As the latter increases, the spectrum starts to resemble that of a chiral fermion, hinting that the modular Hamiltonian approaches that of the 1D conformal field theory (CFT) on the edge of region $A$ \cite{PhysRevLett.96.110404,Li_2008,2012PhRvB..85k5321D}.  We return to this in Sec.\ \ref{seFQHE}, where the Li-Haldane conjecture will be used to estimate symmetry-resolved EE in fractional quantum Hall states.

\subsection{Charge fluctuation operator}
\label{sechop}

The entanglement spectrum of Fig.\ \ref{fispec} contains the full information needed to deduce symmetry-resolved entropies. One may thus factorize the RDM in sectors \eqref{e3} with fixed numbers of particles, and investigate the probabilities $p_q$ and symmetry-resolved density matrices $\rhom(q)$. For convenience, we let the U(1) charge measure fluctuations away from the average:
\be
\label{e35}
\Q_A
\equiv
\nord{\N_A}
-\,\langle\nord{\N_A}\rangle\,,
\ee
where $\langle\cdot\rangle\equiv\text{Tr}(\rhom\,\cdot)$ and normal ordering is required because $\langle \N_A\rangle$ is infinite. As a consequence, the allowed values $q$ of the charge in region $A$ depend on the flux $\Phi$ and need not be integers. Indeed, owing to  $\langle\nord{\cc_m^{\dagger}\cc_m}\rangle=\lambda_m$ for $m\geq0$ and $\langle\nord{\cc_m^{\dagger}\cc_m}\rangle=\lambda_m-1$ for $m<0$, one has
\be
\label{nexp}
\langle\nord{\N_A}\rangle
=
\sum_{m=0}^{+\infty}\lambda_m-\sum_{m=-\infty}^{-1}(1-\lambda_m)
\ee
which may take any real value (\eg $\langle\nord{\N_A}\rangle = 1/2$ for $\Phi =0$). Writing the overlaps \eqref{ovlap} in terms of an error function, one explicitly finds \footnote{To obtain \eqref{e37}, differentiate \eqref{nexp} with respect to $\Phi$, apply Poisson resummation and integrate back over $\Phi$.}
\be
\label{e37}
\langle\nord{\N_A}\rangle
=
\frac{1}{2}
-\Phi
-\sum_{n=1}^{\infty}
\frac{\e^{-\frac{n^2}{4}L^2}}{\pi n}
\sin (2\pi n\,\Phi)
\equiv
-\delta(\Phi)\,.
\ee
This reduces to $\langle\nord{\N_A}\rangle=\frac{1}{2}-\Phi+\cO\big(\e^{-L^2/4}\big)$ in the thermodynamic limit $L\to\infty$. The eigenvalues $q$ of the normalized charge \eqref{e35} thus belong to the set $\ZZ+\delta(\Phi)$; those are the values that will eventually appear in the decomposition \eqref{e3} of the RDM. In particular, $q$ takes half-integer values at zero flux, and integer values for $\Phi=1/2$. 

The interpretation of \eqref{e37} becomes clear upon noting that any change of flux $\Phi$ amounts to a translation of region $A$ along $x$. If the system were invariant under $x$ translations, one would have $\langle\nord{\N_A}\rangle=\frac{1}{2}-\Phi$ without corrections; this occurs in the strict limit $L=\infty$, where the filled LLL becomes translation-invariant. By contrast, at finite $L$, the magnetic field breaks translation symmetry along the cylinder: while translations along $x$ are classical symmetries (they preserve the magnetic field and the metric), they cannot be lifted to the prequantum line bundle because they are not Hamiltonian diffeomorphisms \cite{bates1997lectures}. Aside from \eqref{e37}, this lack of translation invariance is manifest in the exponentially small spatial modulation of the particle density $\rho(\bbr)=\CC(\bbr,\bbr)$ that can be read off from \eqref{e20} with occupied states \eqref{phim}.

\subsection{Exact charged moments}
\label{sefofree}

As recalled in Sec.\ \ref{seSymmRes}, symmetry-resolved entropies are Fourier transforms of charged moments \eqref{zefour}. Using the normal-ordered Hamiltonian \eqref{e32} in the RDM \eqref{e18} then shows that the charged moment \eqref{zefour} reads
\be
\begin{split}
\widehat{Z}_n(\alpha)
=&
\prod_{m<0}
\frac{\text{tr}\left(\e^{i\alpha(1-\lambda_m-\cc_m\cc_m^{\dagger})}\,\e^{n\epsilon_m\cc_m\cc_m^{\dagger}}\right)}
{(1+\e^{\epsilon_m})^n}\\
&
\prod_{m\geq0}
\frac{\text{tr}\left(\e^{i\alpha(\cc_m^{\dagger}\cc_m-\lambda_m)}\,\e^{-n\epsilon_m\cc_m^{\dagger}\cc_m}\right)}
{(1+\e^{-\epsilon_m})^n}\,,
\end{split}
\ee
where the $\lambda_m$'s are overlaps \eqref{ovlap}, pseudo-energies $\epsilon_m$ are given by \eqref{em}, and each lowercase trace is taken over a single fermionic two-level system produced by $\cc_m,\cc_m^{\dagger}$. Evaluating such traces yields
\be
\label{zn}
\widehat{Z}_n(\alpha)
=
\prod_{m\in\ZZ}
\left(\lambda_m^n\e^{i\alpha(1-\lambda_m)}+(1-\lambda_m)^n\e^{-i\alpha \lambda_m}\right)\,.
\ee
Expressions of this kind are well known in the literature (see \eg \cite{Bonsignori_2019}). The only modification here is a slight rearrangement due to normal-ordering, leading to manifestly finite expressions since $\lambda_m^n\e^{i\alpha(1-\lambda_m)}+(1-\lambda_m)^n\e^{-i\alpha \lambda_m}\to1$ exponentially fast when $m\to\pm\infty$. This, so far exact, equation will be our starting point in Sec.\ \ref{seSRE} to evaluate FCS and symmetry-resolved R\'enyi entropies. 

Incidentally, the failure of $q$ to take integer values means that charged moments are {\it not} $2\pi$-periodic in $\alpha$. They are rather quasi-periodic, since $\widehat{Z}_n(\alpha+2\pi)=\e^{2\pi i\delta(\Phi)}\widehat{Z}_n(\alpha)$ as anticipated below \eqref{zefour}. Also note that \eqref{zn} enjoys the relation $\widehat{Z}^{(\Phi)}_n(\alpha)=\widehat{Z}^{(-\Phi)}_n(-\alpha)$, reflecting symmetry under the exchange of regions $A$ and $B$ coupled to $\Phi\mapsto-\Phi$. This invariance is manifest from the relation $\lambda(-k)=1-\lambda(k)$ for overlaps \eqref{ovlap}. For symmetry-resolved entropy, it implies $S^{(\Phi)}_n(q)=S^{(-\Phi)}_n(-q)$, so $S_n(q)$ is even in $q$ when $\Phi$ is an integer or a half-integer. This symmetry is manifest in Figs.\ \ref{firenyi}--\ref{fivn} and Fig.\ \ref{fig:laughlinhalfentropy} below.

\section{FCS and entropies}
\label{seSRE}

Section \ref{sePREL} provides the tools needed to compute sym\-met\-ry-\-resolved measures of entanglement in the thermodynamic limit $L\gg1$. Accordingly, we now evaluate charged moments \eqref{zefour} at large $L$, then integrate over $\alpha$ to read off the probabilities \eqref{fcs}, and finally take logarithms to obtain R\'enyi entropies whose derivative \eqref{e12} yields von Neumann entropy. As we shall see, fluctuations satisfy a Gaussian distribution with deviation $\sqrt{L}$ at leading order (see \eqref{e54}), leading to entropies \eqref{eq:Sqcorrections} that become evenly distributed among different charge sectors in the thermodynamic limit \cite{Laflorencie_2014,PhysRevB.98.041106}. Equipartition of entropy is thus reproduced up to charge-dependent corrections of order $\cO(1/L)$ that we derive explicitly. Similar corrections will eventually affect the fractional QHE in Sec.\ \ref{seFQHE}.

\subsection{Charged moments at large \texorpdfstring{$L$}{L}}
\label{sefoufou}

\begin{figure}[t]
\centering
\includegraphics[width=.79\columnwidth]{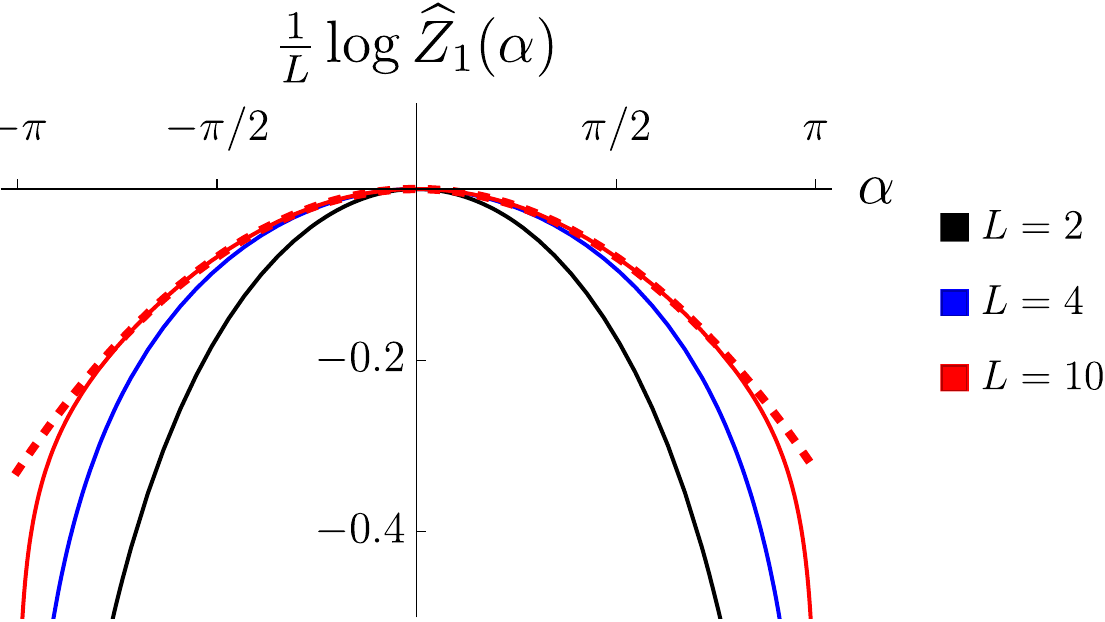}
\caption{Comparison between the exact expression \eqref{zn} for $\widehat{Z}_1(\alpha)$ with $\Phi=0$ for perimeters $L=2,4,10$, and its (dashed) integral approximation \eqref{e51}. Both $\log\widehat{Z}_1(\alpha)$ and its approximation are divided by $L$ for readability. The approximation becomes more accurate as $L$ grows, save for the divergence at $\alpha=\pm\pi$ (corresponding to $\widehat{Z}_1(\pm\pi)=0$ when $\Phi=0$) that cannot be captured by the integral \eqref{e51}. Around its maximum at $\alpha=0$, $\log \widehat{Z}_1(\alpha) $ behaves as a concave parabola, eventually giving the near-Gaussian behavior \eqref{e52}. See also Fig.\ \ref{fizgo}.}
\label{filoz}
\end{figure}

Consider as before a filled LLL of non-interacting electrons, with the region $A$ introduced in Sec.\ \ref{seLLL}. The Fourier transform of FCS is the charged moment \eqref{zn} with $n=1$ and overlaps $\lambda_m$ given by \eqref{ovlap}; our task is to find its large $L$ asymptotics. We shall do this for arbitrary $n$ to streamline the presentation, as charged moments with $n>1$ will be needed for R\'enyi entropies. Accordingly, rewrite the product \eqref{zn} as the exponential of a sum of logarithms and convert the sum over $m$ into an integral at large $L$ (Euler-Maclaurin approximation). Up to exponentially small corrections as $L \to \infty$, the charged moment for $\alpha\in\,]-\pi,\pi[$ becomes
\be
\label{e51}
\!\!\!\!\!\widehat{Z}_n(\alpha)
\sim
\exp\!\left[%
\!\frac{L}{4\pi}\!\!
\int\!\!\dd k
\log\!\Big(
\lambda^{2n}+\bar\lambda^{2n}+2\lambda^n\bar\lambda^n\cos\alpha
\Big)\!
\right]\!\!\!\!\!
\ee
with $\bar\lambda(k)\equiv1-\lambda(k)\equiv\lambda(-k)\equiv\tfrac{1}{2}\erfc(-k)$ as in \eqref{ovlap} and the integral over $k$ runs over the whole real line. For $n=1$, this implies that all even cumulants of charge fluctuations are $\cO(L)$ while odd cumulants are exponentially small, in accordance with \cite{2019CMaPh.376..521C}. More generally $\widehat{Z}_n(\alpha)$ is real, positive and $\Phi$-independent at leading order on the interval $]-\pi,\pi[$, with a maximum at $\alpha=0$. This is {\it not} true for other ranges of $\alpha$, as the integrand of the Euler-Maclaurin formula becomes discontinuous and a non-zero phase appears, consistently with $\widehat{Z}_n(\alpha+ 2\pi)=\e^{2\pi i\delta(\Phi)}\widehat{Z}_n(\alpha)$. Furthermore, the asymptotic relation \eqref{e51} is {\it false} when $\Phi$ is an integer and $\alpha=\pi$ mod $2\pi$, since one then has $\widehat{Z}_n(\pi)=0$ owing to \eqref{zn}. See Fig.\ \ref{filoz} for a comparison between the exact function $\widehat{Z}_1(\alpha)$ and its integral approximation \eqref{e51}.

\begin{figure}[t]
\centering
\includegraphics[width=0.79\columnwidth]{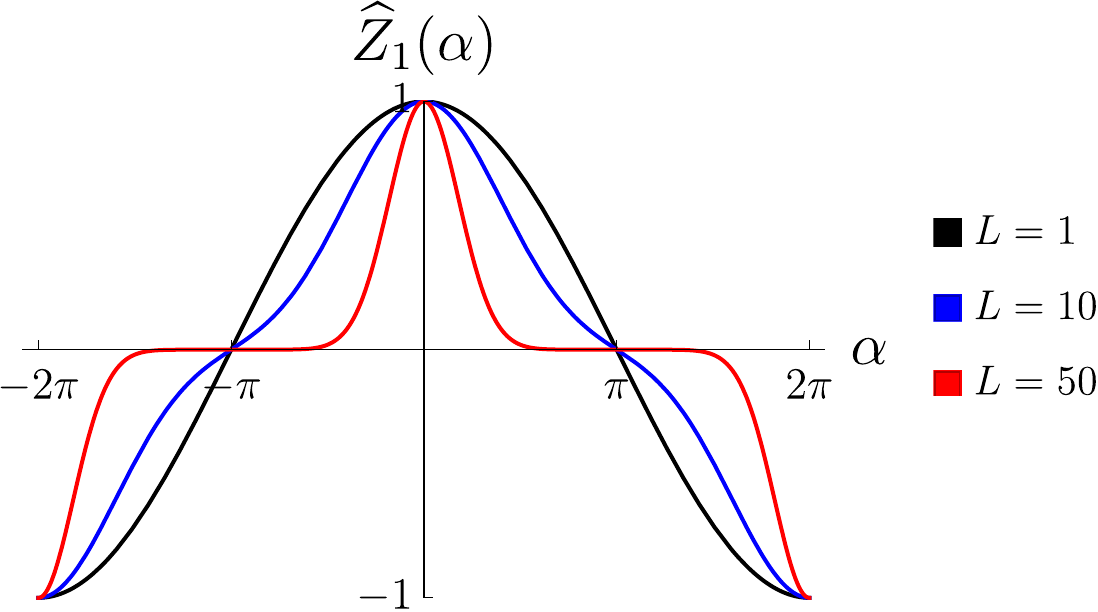}
\caption{The charged moment \eqref{zn} for $n=1$, $\Phi=0$, $\alpha\in[-2\pi,2\pi]$ and $L=1,10,50$. The Gaussian behavior \eqref{e52} is manifest at large $L$, as is the presence of similar Gaussian minima at $\alpha=2\pi$ mod $4\pi$. Note that $\widehat{Z}_1(\alpha)$ is real and antiperiodic when $\Phi=0$, since \eqref{e37} yields $\delta(0)=1/2$.}
\label{fizgo}
\end{figure}

Crucially, only the behavior for $\alpha\in\,]-\pi,\pi[$ is needed for the Fourier transform \eqref{zen}. In fact, at large $L$, $\widehat{Z}_n(\alpha)$ behaves as a sharply peaked Gaussian at $\alpha=0$ (see Fig.\ \ref{fizgo}): Taylor-expanding the exponent of \eqref{e51} yields
\be
\label{e52}
\widehat{Z}_n(\alpha)
\sim
\exp\Big[-L\big(a_n+b_n\alpha^2+c_n\alpha^4\big)+\cO(L\alpha^6)\Big]
\ee
as $L\to\infty$ and $\alpha\to0$, with $L$-independent coefficients
\begin{align}
a_n
&\equiv
-\int\frac{\dd k}{2\pi}
\log(\lambda^n+\bar\lambda^n)\,,\nonumber\\
\label{bn}
b_n
&\equiv
\int\frac{\dd k}{4\pi}
\frac{\lambda^n\bar\lambda^n}{(\lambda^n+\bar\lambda^n)^2}\,,\\
c_n
&\equiv
\int\frac{\dd k}{8\pi}
\left(%
\frac{\lambda^{2n}\bar\lambda^{2n}}{(\lambda^n+\bar\lambda^n)^4}
-\frac{\lambda^n\bar\lambda^n/6}{(\lambda^n+\bar\lambda^n)^2}
\right)\,.\nonumber
\end{align}
Here $a_1=0$ since $\widehat{Z}_1(\alpha)$ is the Fourier transform of a probability distribution; all other $a_n,b_n,c_n$'s are strictly positive. Note for future reference that these coefficients satisfy the large $n$ relations $a_n=n\tilde f-\pi^{3/2}/(48n)+\cO(1/n^3)$ with $\tilde f\simeq0.107483$, $b_n=1/(16\sqrt{\pi}\,n)+\cO(1/n^3)$, and $c_n=\cO(1/n^3)$ \footnote{The proof of the large $n$ asymptotics of coefficients \eqref{bn} follows from their definition as integrals of error functions. In all three cases, the integrand is localized near $k=0$ at large $n$ and the $\cO(1)$ variable that needs to be integrated over is $nk$, which yields consistent expansion schemes in the limit $n\to\infty$.}. We return to this behavior at the end of Sec.~\ref{sec:corrections}, where it will provide a striking check of the Li-Haldane conjecture.

The expansion \eqref{e52} turns out to be crucial. We will soon use it to evaluate FCS and symmetry-resolved entropy for non-interacting electrons, but essentially the {\it same} derivation will apply to the fractional QHE in Sec.~\ref{sec:corrections}. Indeed, \eqref{e52} displays a universal scaling $\widehat{Z}_n(\alpha)\sim\e^{-Lf(\alpha)}$ that also holds in interacting setups, up to Taylor coefficients of $f(\alpha)$ not being given by \eqref{bn}.

\subsection{Full counting statistics}
\label{sec:fullcountingIQHE}

\begin{figure}[t]
\centering
\includegraphics[width=.49\columnwidth]{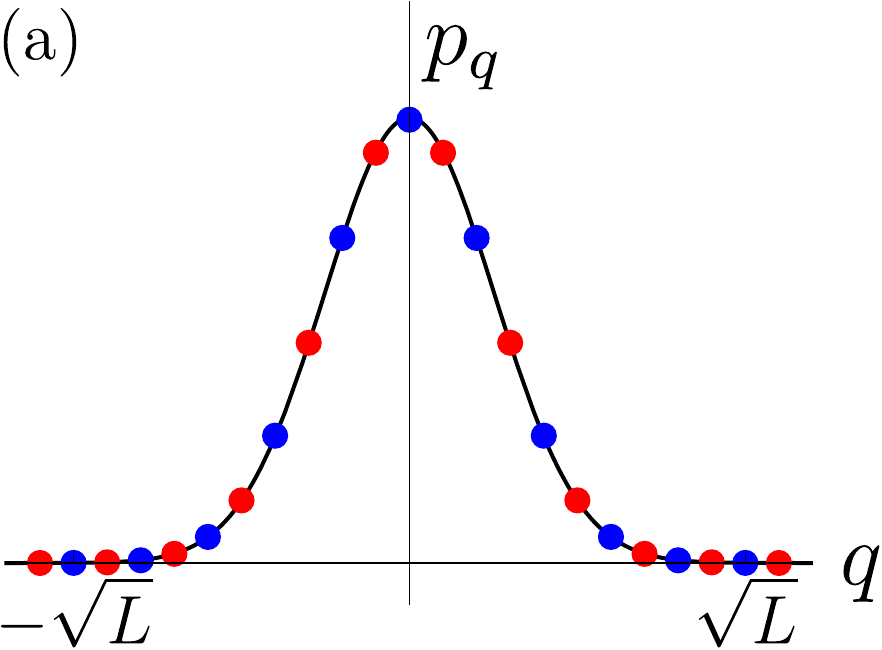}
\hfill
\includegraphics[width=.49\columnwidth]{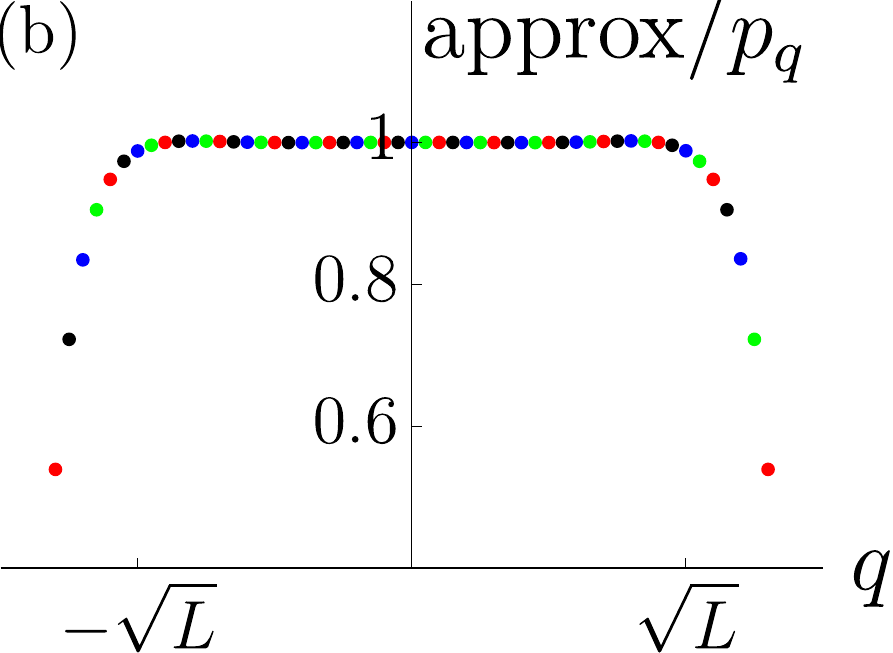}
\caption{FCS (a) on a cylinder with perimeter $L=25$ and its comparison (b) with the approximate nearly Gaussian distribution \eqref{e54}. The dots are exact values given by \eqref{zen}--\eqref{zn} with $n=1$ and fluxes $\Phi=0$ (red), $\Phi=0.25$ (green), $\Phi=0.5$ (blue), $\Phi=0.75$ (black). The approximation \eqref{e54} seems indistinguishable from the exact $p_q$ but it is unreliable for large fluctuations, where $p_q$ drops to zero exponentially fast.}
\label{fisuccess}
\end{figure}

The charged moment \eqref{zn} and its approximations \eqref{e51}--\eqref{e52} provide direct access to the probability distribution \eqref{fcs} of charge fluctuations. Indeed, assuming $q=\cO(\sqrt{L})$, the Fourier transform \eqref{zen} applied to the exponential \eqref{e52} readily yields the large $L$ expansion
\be
\label{e54}
\begin{split}
&Z_n(q)
\sim
\frac{1}{\sqrt{4\pi Lb_n}}
\,\e^{-La_n}
\,\e^{-q^2/(4Lb_n)}\,\times\\
&\times\left[%
1
-\tfrac{c_n}{Lb_n^2}\left(%
\tfrac{3}{4}
-\tfrac{3q^2}{4Lb_n}
+\tfrac{q^4}{16L^2b_n^2}
\right)
+\cO(1/L^2)
\right]\,.
\end{split}
\ee
Since $a_1=0$ and $b_1=(32\pi^3)^{-1/2}$, it follows that leading-order FCS is a Gaussian whose variance is proportional to the perimeter \cite{2006JPhA...39L..85K}:
\be
\label{eq:fcsGauss}
p_q
\sim
\frac{1}{\sqrt{2\pi\sigma^2}}
\e^{-\frac{q^2}{2\sigma^2}}\,,
\qquad
\sigma^2
=
\frac{L}{(2\pi)^{\frac{3}{2}}}\,.
\ee
(See Fig.\ \ref{fisuccess}.) Note that this holds regardless of the flux: in the thermodynamic limit, $\Phi$ only affects FCS through the allowed values $q\in\ZZ+\delta$ of charge fluctuations (recall \eqref{e37}), otherwise leaving the distribution untouched.

We stress that \eqref{e54} is valid for fluctuations $q=\cO(\sqrt{L})$, meaning that terms of the form $q^2/L$ are really of order one. One can indeed verify that the distribution \eqref{e54} is properly normalized for $n=1$ (up to and including order $1/L$), precisely owing to the combination of constant, quadratic and quartic terms in $q$ that appear in the $1/L$ correction. In fact, \eqref{e54} displays a recurring pattern that will also affect entropies, both in the integer QHE and in its fractional version: symmetry-resolved quantities have a $q$-independent dominant term at large $L$, followed by an $\cO(1)$ correction involving $q^2/L$, followed by an $\cO(1/L)$ correction involving both $q^2/L$ and $q^4/L^2$, and so on. Large $L$ expansions thus become tied to polynomial approximations in $q^2/L$. We will rely on this insight in Sec.\ \ref{sec:numericalresults} to choose `educated fits' for symmetry-resolved entropies of fractional quantum Hall states. 

\subsection{Symmetry-resolved entropies}
\label{sec:symresent}

\begin{figure}[t]
\centering
\includegraphics[width=0.79\columnwidth]{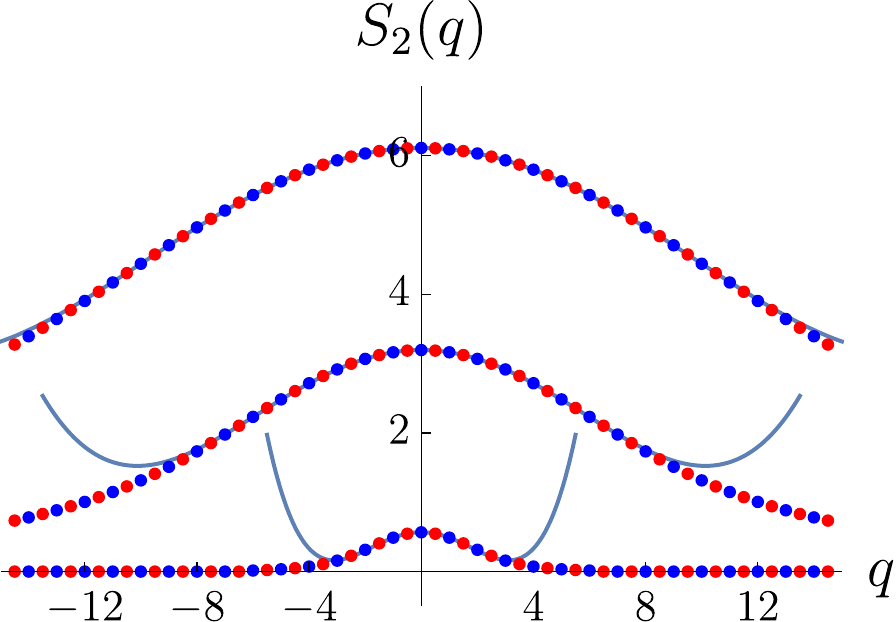}
\caption{Second R\'enyi entropy per sector on cylinders with perimeter (bottom to top) $L=10$, $L=30$, $L=50$ and flux $\Phi=0$ (red), $\Phi=0.5$ (blue). In each case, the solid line is the approximation \eqref{eq:Sqcorrections} that includes constant, quadratic, and quartic terms in $q$ with coefficients \eqref{s8b}. The matching is already fair for relatively small systems ($L=10$); it improves for larger ones. The area law is also visible (the maximum increases linearly with $L$), as is equipartition at large $L$.}
\label{firenyi}
\end{figure}

The large $L$ expansion \eqref{e54} can readily be used in \eqref{fcs} to derive R\'enyi entropies. As a function of $q^2/L=\cO(1)$, the $n^{\text{th}}$ symmetry-resolved R\'enyi entropy thus reads
\begin{empheq}[box=\othermathbox]{equation}
\label{eq:Sqcorrections}
S_n(q)
\sim
S_n
-\tfrac{1}{2}\log L
+A_n
-B_n\frac{q^2}{L}
+C_n\frac{q^4}{L^3}
\end{empheq}
up to corrections of order $\cO(1/L^2)$. Here $S_n=La_n/(n-1)$ is the total R\'enyi entropy and $A_n,B_n,C_n$ are $\cO(1)$ coefficients, positive at large $L$, that each admit their own expansion in powers of $1/L$. Namely
\begin{align}
A_n
&\sim
\tfrac{\log b_n-n\log b_1}{2(n-1)}
-\tfrac{\log(4\pi)}{2}
-\tfrac{3(nc_1/b_1^2-c_n/b_n^2)}{4(n-1)L}\,,\nonumber\\
\label{s8b}
B_n
&\sim
\tfrac{n/b_1-1/b_n}{4(n-1)}
-\tfrac{3(nc_1/b_1^3-c_n/b_n^3)}{4(n-1)L}\,,\\
C_n
&\sim
\tfrac{c_n/b_n^4-nc_1/b_1^4}{16(n-1)}\nonumber
\end{align}
for $n>1$, in terms of constants $a_n,b_n,c_n$ defined in \eqref{bn} and up to corrections of respective order $\cO(1/L^2)$, $\cO(1/L^2)$, $\cO(1/L)$. The behavior of symmetry-resolved von Neumann entropy \eqref{e12} is similar: \eqref{eq:Sqcorrections} remains valid for $n=1$, up to the fact that coefficients $A_1,B_1,C_1$ are not given by \eqref{s8b} but stem instead from the derivative in \eqref{e12}. A straightforward computation based on the integrals \eqref{bn} then yields values of $A_1,B_1,C_1$ that are relegated to Appendix \ref{app:asymptotics} for brevity.

\begin{figure}[t]
\centering
\includegraphics[width=0.79\columnwidth]{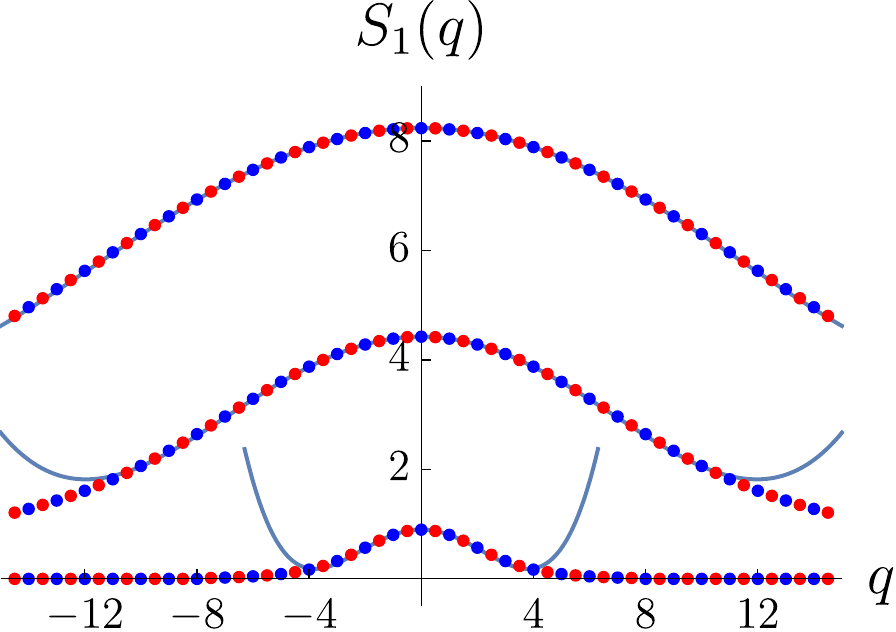}
\caption{Symmetry-resolved EE on cylinders with perimeter (bottom to top) $L=10$, $L=30$, $L=50$ and flux $\Phi=0$ (red), $\Phi=0.5$ (blue). As in Fig.\ \ref{firenyi}, the solid line is the quartic approximation \eqref{eq:Sqcorrections}, now with coefficients $A_1,B_1,C_1$ whose values are written in Appendix \ref{app:asymptotics}. The area law and equipartition at large $L$ are once again manifest.}
\label{fivn}
\end{figure}

The expansion \eqref{eq:Sqcorrections} highlights different aspects of symmetry-resolved EE: the leading term $S_n(q)\sim S_n$ is proportional to $L$, exhibiting the usual area law. Its $\cO(\log L)$ correction lowers the entropy per sector with respect to its leading value, reflecting the fact that each charge sector now carries some fraction of the total entropy while preserving equipartition. In the von Neumann case $n=1$, this logarithmic lowering may be interpreted as a contribution of the entropy of charge fluctuations, in accordance with the decomposition \eqref{vndec}. Further charge-dependent corrections occur at order $\cO(1)=\cO(q^2/L)$. However, for `small' fluctuations $q=\cO(1)$, $q$-dependent terms become negligible in the thermodynamic limit $L\to\infty$, leading to an equipartition of entropy \cite{Laflorencie_2014,PhysRevB.98.041106}. Figures \ref{firenyi}--\ref{fivn} display the excellent fit between the approximate expressions \eqref{eq:Sqcorrections} and their exact numerical values for several cylinder perimeters. The same polynomial dependence of entropy on $q^2/L$ will occur in the fractional QHE in Sec.\ \ref{seFQHE}.

\section{Symmetry-resolved entanglement in the fractional Hall effect}
\label{seFQHE}

This section is devoted to FCS and symmetry-resolved EE in Laughlin states. In contrast to Secs.~\ref{sePREL}--\ref{seSRE}, strong interactions now invalidate the use of free fermion methods: to the best of our knowledge, exact results are unavailable. Instead we provide field-theoretic arguments based on the Li-Haldane conjecture \cite{Li_2008} to show that equipartition still holds for fractional quantum Hall states in the thermodynamic limit $L \gg 1$, up to corrections of order $\cO(q^2/L)$. This is supported by strong numerical evidence based on matrix product state (MPS) simulations for the $\nu=1/2$ Laughlin wavefunction.

\subsection{Bulk-edge correspondence}
\label{sec:conformalapproximation}

In the seminal paper \cite{Li_2008}, Li and Haldane observed that the entanglement spectrum of fractional quantum Hall states is strikingly similar to the energy spectrum of edge modes in the presence of a physical boundary. We already encountered this behavior in Fig.\ \ref{fispec} above for the integer QHE. The ensuing {\it Li-Haldane conjecture} is a bulk-edge correspondence for entanglement: it states that the modular Hamiltonian is an effective gapless local Hamiltonian on the 1D boundary of region $A$. In fact, hints of this relation already appeared in \cite{PhysRevLett.96.110404}, where it was used to derive topological EE. We now show that the same assumption implies a strict equipartition of EE. Corrections to equipartition are studied in Sec.~\ref{sec:corrections}, and their presence is confirmed numerically in Sec.~\ref{sec:numericalresults}.

The most naive expectation for quantum Hall states (both integer and fractional) is that the modular Hamiltonian equals the conformal edge Hamiltonian:
\be
\label{eq_Li_Haldane_naive}
\hamm
=
\frac{2\pi v}{L}
\Big(L_0-\frac{c}{24}\Big)\,.
\ee
Here $L$ is the length of the boundary of $A$, $L_0$ is the zero-mode of the stress tensor, $c$ is the central charge of edge modes and $v$ is a non-universal velocity. For instance, the edge Hamiltonian of a $\nu=1/p$ Laughlin state is $L_0=\tfrac{1}{2} a_0^2+\sum_{n>0}a_{-n}a_n$ in terms of the bosonic modes $a_n$ of a U(1) current. In particular, the electric charge in region $A$ is measured by the zero-mode $a_0=\sqrt{p}\,Q_A$, which takes values in $\ZZ/\sqrt{p}$. (We assume for simplicity that $\Phi = 1/2$; otherwise one must add a shift $\delta(\Phi)$, as in the integer QHE.) The Hilbert space consists of $p$ topological sectors, each having a definite value of $\Q_A=a/p$ mod $1$ with $a=\{0,1,...,p-1\}$. According to the Li-Haldane conjecture in the conformal approximation \eqref{eq_Li_Haldane_naive}, the RDM in sector $a$ reads
\be
\label{eq_Li_Haldane_Laughlin}
\rhom
=
\frac{1}{Z_a}
\e^{-\frac{\pi vp}{L}\Q_A^2}
\e^{-\frac{2\pi v}{L}
\left(\sum_n a_{-n} a_n-1/24\right)}\,,
\ee
where the normalization $Z_a$ is an unspecialized character that can be written in terms of theta functions, but whose exact form is unimportant for our purposes. The RDM \eqref{eq_Li_Haldane_Laughlin} is manifestly block-diagonal with respect to $\Q_A$ in the sense of \eqref{e3}. Even more remarkably, its normalized blocks $\rhom(q)$ do not depend on $q$, so equipartition of entanglement is guaranteed: 
\be
\rhom(q)
=
\frac{1}{\eta(\tau)}
\e^{-\frac{2\pi v}{L}\left(\sum_n  a_{-n}  a_n -1/24\right)}\,.
\ee
where $\eta(\tau)$ is the Dedekind eta function. The corresponding probability of having $\Q_A =q$ is Gaussian, with a normalization that generally depends on the anyonic sector at finite $L$. This dependence goes away at large $L$ and FCS simply becomes
\be
\label{eq:variance}
p_q
\sim
\frac{\e^{-\frac{q^2}{2 \sigma^2}}}{\sqrt{2\pi\sigma^2}}\,,
\qquad
\sigma^2
=
\frac{L}{2\pi p v}\,,
\qquad
q\in \frac{a}{p} +\ZZ\,.
\ee
This generalizes the Gaussian \eqref{eq:fcsGauss} of the integer QHE.

\subsection{Corrections to the conformal spectrum}
\label{sec:corrections}

While rather compelling, the above analysis rests on a strong assumption: that the modular Hamiltonian is strictly conformal. The actual situation is neither so simple, nor universal: finite size effects imply that \eqref{eq_Li_Haldane_naive} does not hold exactly, as the entanglement spectrum is not strictly linear. For the integer QHE, this is manifest in Figs. \ref{fidispersion}--\ref{fispec} above. One of the profound insights of \cite{2012PhRvB..86x5310D} is that the finite size entanglement spectrum can be recovered by adding irrelevant, local perturbations to the CFT Hamiltonian \eqref{eq_Li_Haldane_naive}:
\be
\label{eq_Li_Haldane_perturbed}
\hamm
=
\frac{2\pi v}{L}  \left(  L_0  - \frac{c}{24} \right) + \sum_{j}g_{j} 
\underbrace{\int_0^L \phi_j(y) \,\dd y}_{\equiv(\tfrac{\pi}{L})^{\Delta_j-1} \,\displaystyle V_j}
\ee
where $\phi_j$'s are local fields with scaling dimensions $\Delta_j > 2$, and $V_j$ stands for the zero-mode of $\phi_j$. On the cylinder geometry we have adopted, symmetry under exchange of regions $A$ and $B$ requires the $\phi_j$'s to be U(1) neutral, leading to $\Delta_j \geq 4$ \cite{2012PhRvB..86x5310D}. For the fractional QHE, such finite size corrections mean that quantitative tests of the Li-Haldane conjecture are challenging and require fine-tuning all the $g_j$'s (see \cite[Appendix E]{2012PhRvB..86x5310D} and the more recent \cite{PhysRevB.104.195434}). Indeed, these coupling constants are not universal: they depend not only on microscopic details, but also on the shape of the boundary of $A$.

In the integer QHE with the geometry used in this paper, the corrections can in fact be computed exactly. Upon normal-ordering the quadratic modular Hamiltonian \eqref{hamm_diagonal} and expanding pseudo-energies as $\epsilon(k)= \sum_{j \geq 0} g_j k^{2j+1}$, one finds an expression of the form \eqref{eq_Li_Haldane_perturbed} with $ V_j\propto\sum_mm^{2j+1}\,:\cc^{\dag}_m\cc_m:$. The term $j=0$ is the usual conformal Hamiltonian of a chiral fermion:
\be
L_0
=
\sum_{m\in\ZZ}
m
\,
:\cc^{\dag}_m\cc_m:\,
=
L \int_0^L
:\Psi^{\dag}(y)(-i\partial_y)\Psi(y):\,\dd y
\ee
where the field $\Psi^{\dag}(y)\equiv\sum_m\e^{i\frac{2\pi m}{L}y} \cc^{\dag}_m$ creates electrons at the interface $\partial A$. Each perturbation $ V_j$ for $j \geq 1$ is manifestly local and irrelevant, since it is proportional to the zero-mode of the local field $:\Psi^{\dag}(y) (-i \partial_y)^{2j+1} \Psi(y) :$ with scaling dimension $\Delta_j=2j +2$. 

\begin{figure}[t]
\centering
\includegraphics[width=0.79\columnwidth]{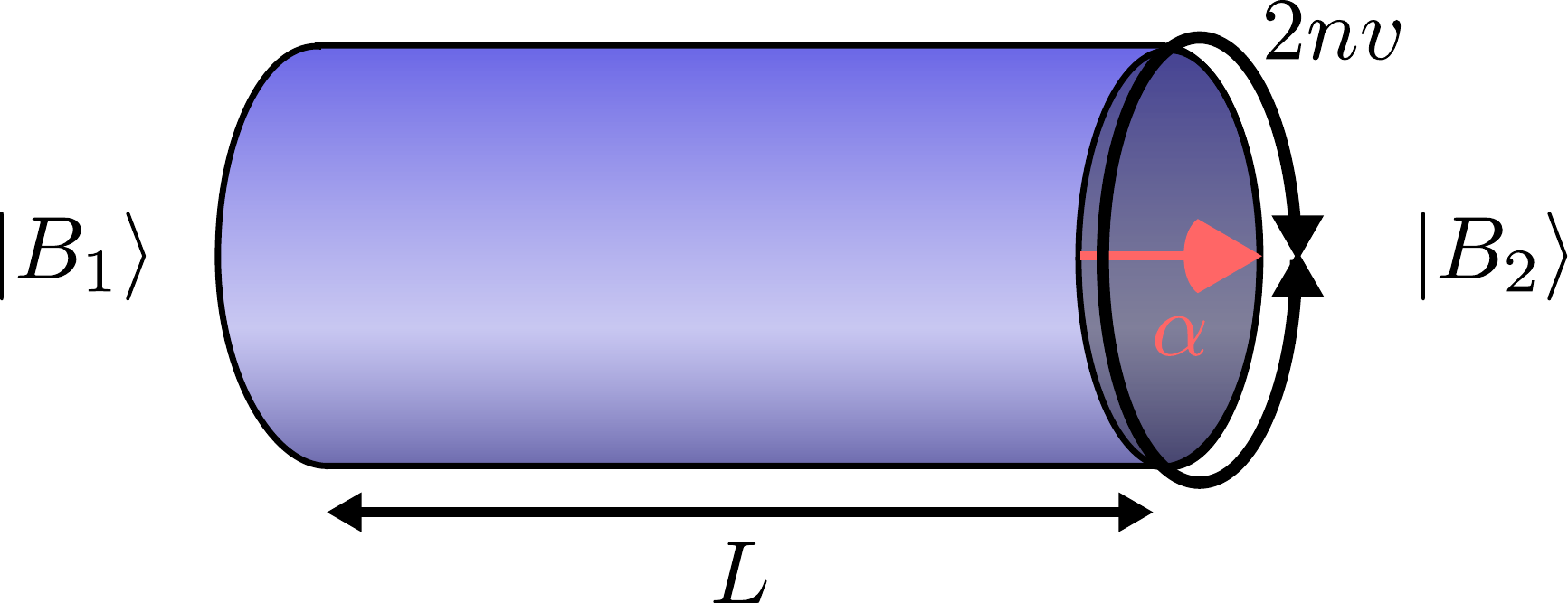}
\caption{The Euclidean space-time involved in the partition function \eqref{charged_moments_CFT} has the topology of an annulus threaded by a magnetic flux $\alpha$. The partition function can be interpreted in two ways: (i) as an open system of size $L$, inverse temperature $\beta_n = 2 n v$ and complex chemical potential $ i \alpha/\beta_n$; (ii) upon interchanging space and imaginary time, as a periodic system of size $2 n v$ with boundary conditions twisted  by the magnetic flux $\alpha$.  From the second perspective, \eqref{charged_moments_CFT} is the probability amplitude to be in the (boundary) state 
$| B_2 \rangle $ starting from $| B_1 \rangle $, after an imaginary time interval $-i L$.}
\label{fig_annulus}
\end{figure}

Let us now return to Laughlin states. In order to understand the effects of irrelevant deformations \eqref{eq_Li_Haldane_perturbed} on FCS and symmetry-resolved EE, it is easier to first consider the charged moments \eqref{zefour}:
\be
\label{charged_moments_CFT}
\widehat{Z}_n(\alpha)
=
\frac{1}{Z_a^n}
\text{Tr}_a
\Big(%
\e^{i \alpha \Q_A}\e^{-\frac{2 \pi v n}{L}(L_0-1/24)+\cdots}
\Big)
\ee
where $\cdots$ stands for the perturbations in \eqref{eq_Li_Haldane_perturbed} and the trace is taken in the topological sector $a$. The trick is to recognize that the numerator is the partition function of a critical 1D system on an {\it open} chain of length $L$ at inverse temperature $\beta_n=2vn$, with a twist $e^{i\alpha\Q_A}$ (see Fig.\ \ref{fig_annulus}). This system is subject to irrelevant and neutral (bulk) perturbations. Openness accounts for the fact that the Hamiltonian is chiral, and the projection on sector $a$ is achieved by imposing appropriate boundary conditions (to be discussed below). The factor $Z_a$ in the denominator is a partition of the same kind, but at temperature $\beta_1^{-1}$ and with zero twist. The behavior of such partition functions at large $L$ follows from the theory of finite size scaling. Interchanging space and imaginary time, one obtains a periodic system of size $2nv$, with periodic boundary conditions twisted by a phase $\e^{i\alpha}$. The partition function in \eqref{charged_moments_CFT} can thus be recast as 
\be
\label{eq_Wick_rotated}
\langle B_2 |
 \e^{-L H_n}
 |B_1\rangle\,,
\ee
where $|B_1\rangle$ and $|B_2\rangle $ are boundary states in the twisted sector $\alpha$. In the absence of boundary perturbations, they would be conformal boundary states, whose exact form can be obtained from \cite{Cappelli_2002} by performing an appropriate spectral flow to account for twisted boundary conditions:
\be
\label{eq_B_state}
|B\rangle
=
\frac{1}{p^{1/4}}
\sum_{a = 1}^p
\e^{\frac{2\pi i a B}{p}}\,  \ket{a + \alpha/2\pi} \rangle \,,
\ee
where the Ishibashi state $\ket{a} \rangle$ is
\be
\label{eq_ishi_state}
 | a \rangle \rangle
 \equiv
 \sum_{q = a \textrm{ mod } p}   \exp \left( \sum_{j =1}^{\infty} \frac{a_{-j} \overline{a}_{-j}}{j} \right)  \left| \frac{q}{\sqrt{p}} \right\rangle
\ee
and $\ket{q/\sqrt{p}}$ is the highest-weight state in the sector $a_0 = \overline{a}_0 = q/ \sqrt{p}$. Choosing $B_2 = B_1 + a$ ensures the projection on the topological sector $a$ in the initial picture. As for the operator $H_n$ in \eqref{eq_Wick_rotated}, it is the Hamiltonian of a (perturbed) conformal periodic system of length $2nv$:
\be
\label{eq:hn}
 H_n
=
\frac{\pi}{n v}
\left( L_0+{\bar{L}}_0-\frac{1}{12}\right)
+
\sum_{j} \tilde{g}_j
\left(\frac{\pi}{nv}\right)^{\Delta_j-1}
{V}_j\,.
\ee
where $\tilde{g}_j = g_j/2v$. Note that the periodic system becomes large in the limit of large R\'enyi index $n$, so irrelevant perturbations are negligible in that regime and the Hamiltonian \eqref{eq:hn} becomes conformal. We return to the large $n$ limit at the end of this subsection.

The rewriting \eqref{eq_Wick_rotated} makes it clear that the large $L$ limit of the partition function $\langle B_2 |\e^{- L H_n } | B_1 \rangle$ is dominated by the lowest energy eigenstate $\ket{\alpha}$ of $H_n$ in the twisted sector $\alpha$. Thus 
\be
\langle B_2 | \e^{- L  H_n } | B_1 \rangle
\sim
\langle B_2|\alpha\rangle\langle\alpha|B_1\rangle\,
\e^{- L E_n(\alpha)}\,,
\ee
where $E_n(\alpha)$ is the energy of  $\ket{\alpha}$. Furthermore, the coefficient $\bra{\alpha} B \rangle$ is universal and thus robust to irrelevant perturbations \cite{PhysRevLett.67.161}, so \eqref{eq_B_state} yields $\bra{\alpha} B \rangle =p^{-\frac{1}{4}}$ for $\alpha \in  ]-\pi,\pi [$. It follows that the charged moments \eqref{charged_moments_CFT} satisfy the large $L$ relation
\be
\label{eq:chargemomlargeL}
\widehat{Z}_n(\alpha)
\sim
p^{\frac{n-1}{2}} \,\e^{-L(E_n(\alpha) - n E_1(0))}\,.
\ee
For $\alpha =0$, this reproduces the expected scaling of total R\'enyi entropy:
\be
S_n
=
\frac{1}{1-n} \log \widehat{Z}_n(0) \sim \frac{E_n(0) - n E_1(0)}{n-1} L - \gamma\,,
\ee
where $\gamma= \log \sqrt{p}$ is the topological EE of the $\nu=1/p$ Laughlin state \cite{PhysRevLett.96.110404}. For non-zero $\alpha$, the irrelevant perturbations \eqref{eq_Li_Haldane_perturbed} affect the form of $E_n(\alpha)$, which is generally unknown. The best one can do is to rely on general principles: symmetry under charge conjugation implies that $E_n(\alpha)$ is even in $\alpha$, and it is somewhat expected that twisting the boundary conditions involves an energy cost, so $E_n(\alpha)$ should have a minimum at $\alpha=0$. The large $L$ limit is thus dominated by the behavior of $E_n(\alpha)$ near $\alpha=0$, which yields
\be
\label{zna}
\widehat{Z}_n(\alpha)
\sim
p^{\frac{n-1}{2}} \,
\e^{-L\left(a_n+b_n\alpha^2+c_n\alpha^4+\cdots\right)}\,.
\ee
This reproduces the expansion \eqref{e52} found in the integer QHE---of course with different coefficients $a_n,b_n,c_n$. FCS and symmetry-resolved entropies therefore behave in the same way as what was described in Secs.~\ref{sec:fullcountingIQHE}--\ref{sec:symresent} for the integer QHE. In particular, the entropy expansion \eqref{eq:Sqcorrections} remains valid with coefficients $A_n,B_n,C_n$ given by \eqref{s8b}. This coincidence is our key analytical conclusion for fractional quantum Hall states, and it will be verified numerically in Sec.~\ref{sec:numericalresults}.

To conclude, let us consider the limit of large R\'enyi indices. As mentioned above, the irrelevant perturbations of \eqref{eq:hn} become arbitrarily small at large $n$, so the eigenvalue $E_n(\alpha)$ satisfies
\be
\label{eq_large_Renyi_1}
E_n(\alpha)
=
nf
+\frac{2\pi}{vn}\left(h_{\alpha}-\frac{1}{24}\right)
+\cO(n^{1-\Delta})\,.
\ee
Here $f$ is some (non-universal) free energy density, $h_{\alpha}\equiv\frac{1}{2 p} \left( \frac{\alpha}{2\pi} \right)^2$ is the conformal dimension of the primary state $\ket{\alpha}$ and $\Delta\geq4$ is the scaling dimension of the least irrelevant perturbation. Plugging \eqref{eq_large_Renyi_1} back in \eqref{eq:chargemomlargeL} and comparing with the large $L$ behavior \eqref{zna} then predicts the large $n$ relations
\begin{align}
a_n
&=
n(f-E_1(0))
-
\frac{\pi}{12vn}
+\cO(1/n^3)\,,\nonumber\\
\label{eq:bnaspt}
b_n
&=
\frac{1}{4\pi vpn}
+\cO(1/n^3)\,, \\
c_n
&=
\cO(1/n^3)\,.\nonumber
\end{align}
This can be compared with the exact results obtained for the integer QHE in Sec.~\ref{sefoufou}: using $p=1$ and $v=4/\sqrt{\pi}$ owing to the linearized dispersion relation \eqref{em}, Eqs.\ \eqref{eq:bnaspt} predict that the coefficients of charged moments satisfy $a_n=n\tilde f-\pi^{3/2}/(48n)+\cO(1/n^3)$ for some constant $\tilde f$, $b_n=1/(16\sqrt{\pi}\,n)+\cO(1/n^3)$ and $c_n =\cO(1/n^3)$. These are indeed the asymptotics quoted below the definitions \eqref{bn}, thus  providing a highly non-trivial check of the field-theoretic arguments presented here.

\subsection{Numerical results}
\label{sec:numericalresults}

A large family of fractional quantum Hall model states on the cylindrical geometry can conveniently be written as exact MPSs \cite{Zaletel2012,Estienne2013}, including spinful wavefunctions \cite{Crepel2018} and states with quasihole \cite{Zaletel2012,Estienne2013,Wu2015} or quasielectron \cite{Kjall2018} excitations. We now use this MPS description to test the equipartition hypothesis in the fractional QHE. More precisely, we consider FCS and symmetry-resolved entropies, and verify that an expansion of the form \eqref{eq:Sqcorrections} still holds despite strong correlations.

We focus on the Laughlin $\nu=1/2$ state since it has the smallest correlation length in the Laughlin series \cite{Crepel2018}. For a pedagogical derivation, technical details and practical implementation of MPSs for the Laughlin $\nu=1/2$ state on a cylinder, we refer to the appendices of Ref.~\onlinecite{Crepel2018}. The MPS auxiliary space for such model states is the Hilbert space of the underlying CFT, and is therefore infinite-dimensional. Fortunately, it can be efficiently truncated by restricting the conformal dimension of CFT states to be lower than some integer $P_{\rm max}$. This truncation caps the number of both momentum and charge sectors in the entanglement spectrum. Chirality of the entanglement spectrum and the Gaussian decay of FCS thus imply that large enough values of $P_{\rm max}$ allow one to faithfully capture a model state at finite cylinder perimeter. (The calculation complexity grows exponentially with $P_{\rm max}$, however.)  Note that throughout this section, we work with a rescaled charge $\tilde{q}\equiv  Q_A/p$ (for a Laughlin $\nu=1/p$ state). It is directly related to the electric charge up to a factor $e/p$, \ie the elementary excitation charge.

\begin{figure}[!h]
\centering
\includegraphics[width=0.99\columnwidth]{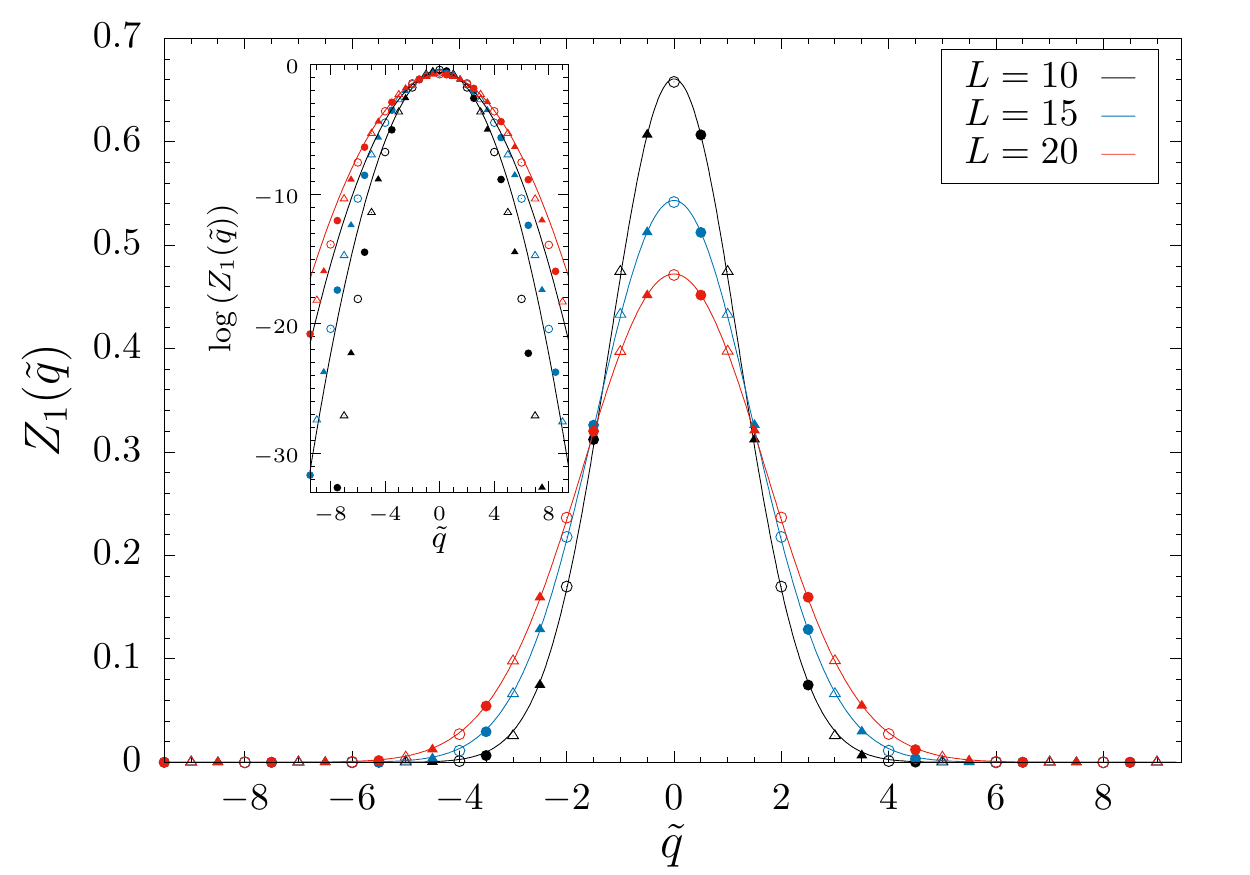}
\caption{FCS for the bosonic Laughlin $\nu=1/2$ state for three perimeters: $L=10$ (black), $L=15$ (blue) and $L=20$ (red). We use two different symbols for the different topological sectors (controlled by boundary conditions), namely circles and triangles. Full symbols indicate zero flux along the cylinder, while hollow symbols stand for a half-flux insertion. The continuous lines are Gaussian fits \eqref{eq:gaussianfitpartition} whose width $\sigma$ is the only fitting parameter. The data were obtained using a truncation $P_{\rm max}=19$.  {\it Inset}: the logarithm of FCS for the same data.}
\label{fig:laughlinhalfchargefluctuations}
\end{figure}

\begin{figure}[!h]
\centering
\includegraphics[width=0.7\columnwidth]{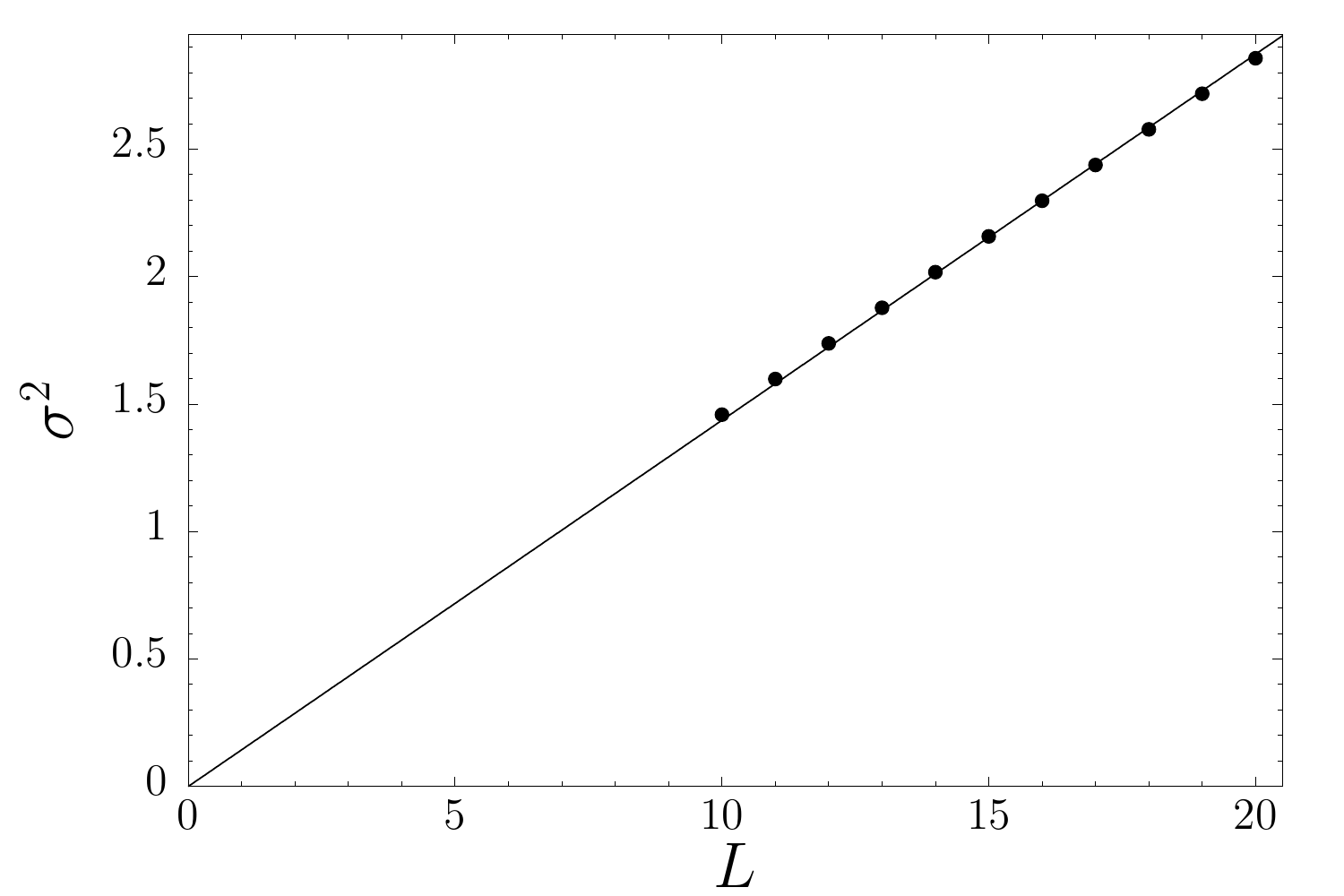}
\caption{Gaussian variance $\sigma^2$ of FCS (as defined in \eqref{eq:gaussianfitpartition}) as a function of the cylinder perimeter $L$ for the Laughlin $\nu=1/2$ state. Error bars are barely visible. The solid line is a linear fit with a slope of $0.144(1)$ whose value measures the velocity $v$ of \eqref{eq:variance}.}
\label{fig:gaussianfit}
\end{figure}

We first focus on FCS on a cylinder with perimeter $L$ between $10$ and $20$, using a truncation $P_{\rm max}=19$. The motivation for this perimeter range and a discussion of CFT truncation effects are provided in Appendix \ref{app:numerics}. Figure \ref{fig:laughlinhalfchargefluctuations} shows the corresponding FCS for different values of $L$. For each size, we combine the two topological sectors of the Laughlin $\nu=1/2$ state. These topological sectors can be thought of as boundary conditions for the infinite cylinder, namely the number of unbalanced excitations at the boundaries modulo $p=2$. As such, each topological sector only gives access to $\tilde{q}$ values with a given parity. Still, both sectors perfectly lay over the same Gaussian distribution \eqref{eq:variance}:
\be
\label{eq:gaussianfitpartition}
p_{\tilde{q}}
=
\frac{p}{\sqrt{2 \pi \sigma^2}}\,
\e^{-\frac{\tilde{q}^2}{2\sigma^2}}\,.
\ee
Note that the prefactor $p$ here stems from the fact that we work with a rescaled charge $\tilde{q}= Q_A/p$. Similarly to what we did for the integer QHE in Sec.~\ref{sec:fullcountingIQHE}, intermediate $\tilde{q}$ values are reached by inserting a magnetic flux along the cylinder. Fitting the numerical data with \eqref{eq:gaussianfitpartition} where $\sigma$ is the only parameter, one can extract the $L$ dependence of the Gaussian width. This is shown in Fig.~\ref{fig:gaussianfit}, exhibiting excellent agreement with the linear behavior of the variance \eqref{eq:variance} derived in Sec.~\ref{sec:conformalapproximation}.

\begin{figure}[!h]
\centering
\includegraphics[width=0.78\columnwidth]{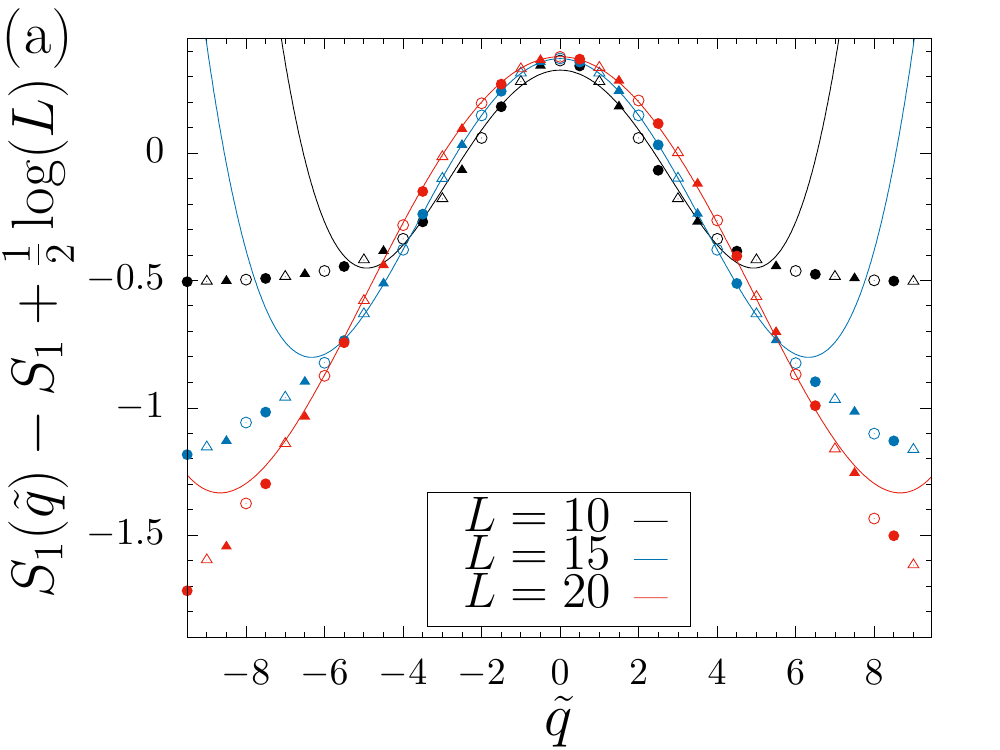}
\includegraphics[width=0.78\columnwidth]{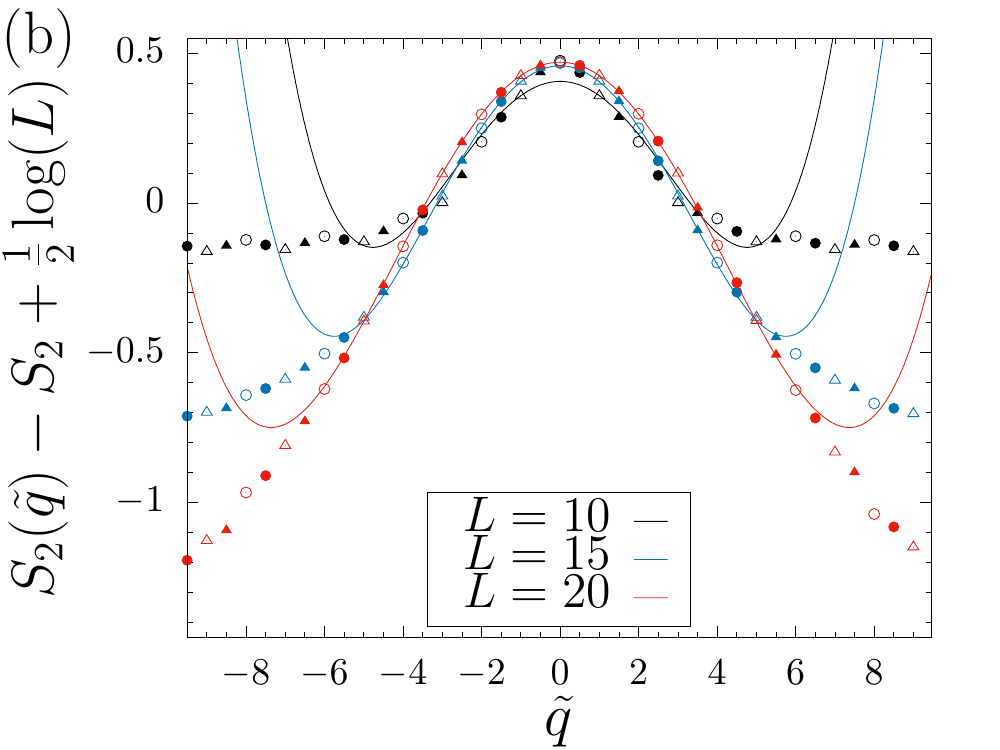}
\caption{Charge-resolved von Neumann (a) and second R\'enyi (b) entropies for the bosonic Laughlin $\nu=1/2$ state for three perimeters: $L=10$ (black), $L=15$ (blue) and $L=20$ (red).  The symbols are identical to those of Fig.~\ref{fig:laughlinhalfchargefluctuations}. The solid lines are quartic fits with three parameters $A_n$, $B_n$, $C_n$ as defined in \eqref{eq:Sqcorrections}, using $|q|<6$ to ensure reliable values (see Appendix \ref{app:numerics}). Similarly to Figs.\ \ref{firenyi}--\ref{fivn}, the quartic approximation \eqref{eq:Sqcorrections} holds for $\tilde{q}=\cO(\sqrt{L})$, so that its range of validity grows with $L$.}
\label{fig:laughlinhalfentropy}
\end{figure}

\begin{figure}[!h]
\centering
\includegraphics[width=0.95\columnwidth]{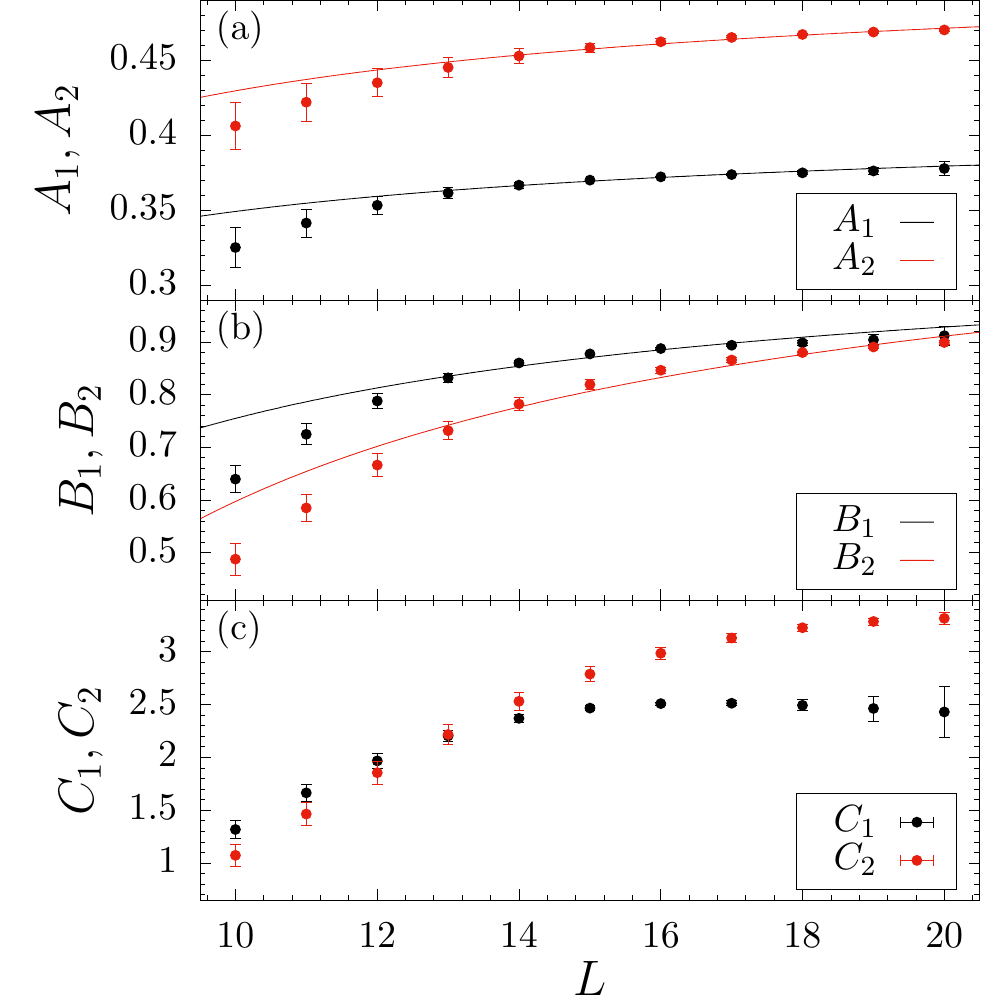}
\caption{Fitting parameters of charge-resolved EE (as defined in \eqref{eq:Sqcorrections}) versus the cylinder perimeter $L$ for the Laughlin $\nu=1/2$ state. From top to bottom: $A_n$ (a), $B_n$ (b) and $C_n$ (c) for both the von Neumann entropy ($n=1$ in black) and the second R\'enyi entropy ($n=2$ in red). For (a) and (b), the solid lines are the $\cO(1)+\cO(1/L)$ fit expected from \eqref{s8b}.}\label{fig:entropyfit}
\end{figure}

We now turn to charge-resolved EE. We consider both von Neumann entropy and the second R\'enyi entropy, the latter usually being less sensitive to finite size effects. Figure \ref{fig:laughlinhalfentropy} then is the analogue of Fig.~\ref{fig:laughlinhalfchargefluctuations} for these two entanglement measures. More precisely, we consider the combination $S_n(\tilde{q}) - S_n + \frac{1}{2} \log (L)$ to directly get access to the $\tilde{q}$-dependent corrections summarized in \eqref{eq:Sqcorrections}, using this expression to fit the data through parameters $A_n$, $B_n$, $C_n$. Despite good overall agreement, some deviations occur due to finite size effects at low perimeter and truncation effects at large perimeter (see also Appendix \ref{app:numerics}). Figure \ref{fig:entropyfit} displays the $L$-dependence of these three parameters, all expected to be of order $\cO(1)$ at large $L$ owing to \eqref{s8b}. For $A_n$ and $B_n$ (Figs.~\ref{fig:laughlinhalfentropy}(a--b)), the agreement is fairly good. Note that error bars increase for small and large perimeters, as expected from the deviations observed in Fig.~\ref{fig:laughlinhalfentropy} and pointed out previously. As for $C_n$, it should be a constant according to \eqref{s8b}---although this is only true up to $1/L$ corrections. The data in Fig.~\ref{fig:laughlinhalfentropy}(c) are indeed compatible with a non-vanishing constant value at large perimeters, but $C_n$ still depends on $L$ over the range that we consider. We emphasize that adding $1/L$ corrections to $C_n$ would also require higher $\tilde{q}^2/L$ terms in \eqref{eq:Sqcorrections} for $\tilde{q}=\cO(\sqrt{L})$. Such terms would be hard to fit with the finite $P_{\rm max}$ we have access to, so we refrain from attempting to verify this finer approximation.

\section{Conclusion}
\label{sec:CON}

This work was devoted to symmetry-resolved entanglement measures in the simplest integer and fractional quantum Hall states. The main goal was to confirm equipartition of entropies in the thermodynamic limit and to derive its least irrelevant charge-dependent corrections. The key result is \eqref{eq:Sqcorrections}: it holds in both free and interacting setups, with coefficients that satisfy specific scaling laws with the length $L$ of the entangling boundary. This ultimately stems from the simple exponential behavior \eqref{e52}--\eqref{zna} of charged moments at large $L$, valid indeed in both integer and fractional QHEs. In the first case, all coefficients can be computed exactly from the known entanglement spectrum of free fermions. In the second, one has to rely instead on field-theoretic arguments such as the bulk-boundary correspondence. As emphasized throughout, the matching between analytical predictions and numerically exact values is excellent. 

It would be interesting to confirm the validity of an expansion of the form \eqref{eq:Sqcorrections} beyond the highly symmetric bipartition considered here, roughly along the lines of what was achieved in \cite{2019CMaPh.376..521C} for (total) EE in arbitrarily-shaped subregions in the integer QHE. A complementary issue also needs to be addressed: our approach systematically consisted in resolving the RDM in sectors with definite electric charge, but it would be equally consistent to resolve it in sectors with definite momentum around the cylinder. Finally, it is natural to wonder how the arguments of Sec.\ \ref{seFQHE} apply to fractional quantum Hall phases supporting {\it non-Abelian} anyons, such as the $\nu=5/2$ state. We hope to address some of these issues in the future.

\section*{Acknowledgements}

We are grateful to Valentin Cr\'epel, J\'er\^ome Dubail and Yacine Ikhlef for fruitful discussions. B.O.\ also acknowledges lively interactions with Gregory Kozyreff on charge fluctuations at large deviations, while B.E.\ thanks Laurent Charles for discussions at an early stage of this work. The work of B.E., N.R.\ and B.O.\ was supported by the ANR grant {\it TopO} No.\ ANR-17-CE30-0013-01. B.E.\ and N.R.\ were also supported by the ANR grant {\it TNStrong} No.\ ANR-16-CE30-0025, while B.O.\ was supported by the European Union's Horizon 2020 research and innovation programme under the Marie Sk{\l}odowska-Curie grant agreement No.\ 846244.

\appendix

\section{Asymptotics of von Neumann entropy}
\label{app:asymptotics}

In Sec.\ \ref{seSRE}, we glossed over the derivation of symmetry-resolved von Neumann entropy since it follows from the derivative formula \eqref{e12} once R\'enyi entropies are expressed in terms of explicit integrals \eqref{bn}. We now rectify this situation and display the values of coefficients $A_1,B_1,C_1$ that appear in the thermodynamic expansion \eqref{eq:Sqcorrections}. This requires one more bit of notation: we define $L$-independent, positive coefficients stemming from the derivatives of \eqref{bn} with respect to $n$ at $n=1$,
\begin{align}
a_0
&\equiv
-\int\frac{\dd k}{2\pi}(\lambda\log\lambda+\bar\lambda\log\bar\lambda)\,,\nonumber\\
b_0
&\equiv
\int\frac{\dd k}{4\pi}
\lambda\bar\lambda
(\log\lambda-\log\bar\lambda)
(\lambda^2-\bar\lambda^2)\,,\\
c_0
&\equiv
\int\frac{\dd k}{4\pi}
\Big(\lambda^2\bar\lambda^2-\frac{\lambda\bar\lambda}{12}\Big)
\Big((2\lambda-1)\log\lambda+(2\bar\lambda-1)\log\bar\lambda\Big)\nonumber
\end{align}
where $\lambda(k)\equiv\tfrac{1}{2}\erfc(k)$ as in \eqref{ovlap} and $\bar\lambda(k)\equiv\lambda(-k)$; all integrals over $k$ cover the entire real line. In these terms, symmetry-resolved von Neumann entropy takes the form \eqref{eq:Sqcorrections} with $n=1$, $S_1=La_0$ and
\begin{align}
A_1
&\sim
\tfrac{\log(4\pi b_1)}{2}
+\tfrac{b_0}{2b_1}
+\tfrac{3(c_0+1/c_1-2b_0c_1/b_1)}{4Lb_1^2}\,,\nonumber\\
B_1
&\sim
\tfrac{1-b_0/b_1}{4b_1}
-\tfrac{3(c_0+c_1-3b_0c_1/b_1)}{4Lb_1^3}\,,\\
C_1
&\sim
\tfrac{4b_0c_1/b_1-c_0-c_1}{16b_1^4}\nonumber
\end{align}
up to neglected corrections of order $\cO(1/L^2)$, $\cO(1/L^2)$, $\cO(1/L)$ respectively. These are the coefficients used in Fig.\ \ref{fivn} to compare exact symmetry-resolved von Neumann entropy to its large $L$ approximation for $q=\cO(\sqrt{L})$.

\section{Additional numerical results}\label{app:numerics}

This appendix provides additional numerical data to motivate our choice of parameters when studying the Laughlin $\nu=1/2$ state in Sec.~\ref{sec:numericalresults}. For that purpose, we consider the topological entanglement entropy (TEE) $\gamma$ \cite{PhysRevLett.96.110404,PhysRevLett.96.110405}, \ie the first correction to the area law that only depends on the nature of excitations:
\be
\label{app:eq:definitiongamma}
S_n
=
\alpha_n L
-\gamma
+\cO(1/L)\,,
\ee
where $\alpha_n$ is some non-universal coefficient. For the Laughlin $\nu=1/p$ states, TEE is given by $\gamma=\log (\sqrt{p})$. Being a subdominant contribution, its extraction provides a simple and reliable proxy to validate a perimeter range for a fixed truncation parameter $P_{\rm max}$ of the MPS calculation. 

\begin{figure}[!h]
\centering
\includegraphics[width=0.99\columnwidth]{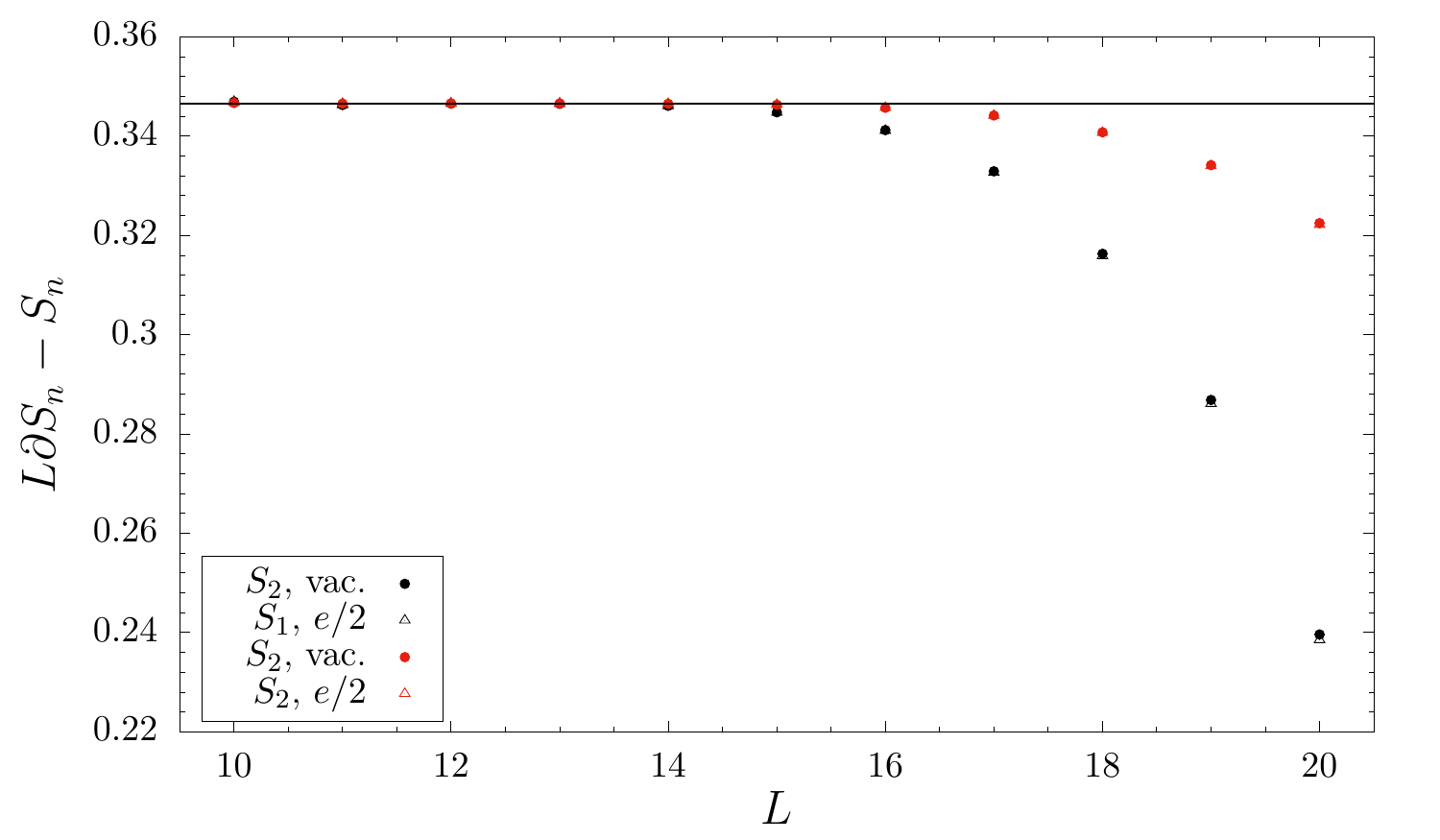}
\caption{TEE for the Laughlin $\nu=1/2$ state extracted from von Neumann EE (black) and the second R\'enyi entropy (red). In practice, we numerically evaluate $L \partial S_n/\partial L - S_n$, whose dominant term is $\gamma$, at different values of the perimeter $L$ \cite{Lauchli_2010}. The horizontal black line is a guide for the eye at the expected value $\gamma=\log (\sqrt{2})$ for the Laughlin $1/2$ state. The numerical derivative is obtained by computing entropies at perimeters $\pm 0.005$ around each integer value of $L$ in the range of interest. The truncation parameter is $P_{\rm max}=19$.}\label{app:fig:entropyderivative}
\end{figure}

Fig.~\ref{app:fig:entropyderivative} shows the extraction of TEE for the Laughlin $\nu=1/2$ state. The value obtained by differentiating EE with respect to $L$ matches the theoretical prediction up to $L\simeq 16$. Beyond that point, deviations appear and are more severe for von Neumann EE than the second R\'enyi EE (as mentioned in Sec.~\ref{sec:numericalresults}). This is why we restrict attention to values $L\leq20$.

\begin{figure}[!h]
\centering
\includegraphics[width=0.99\columnwidth]{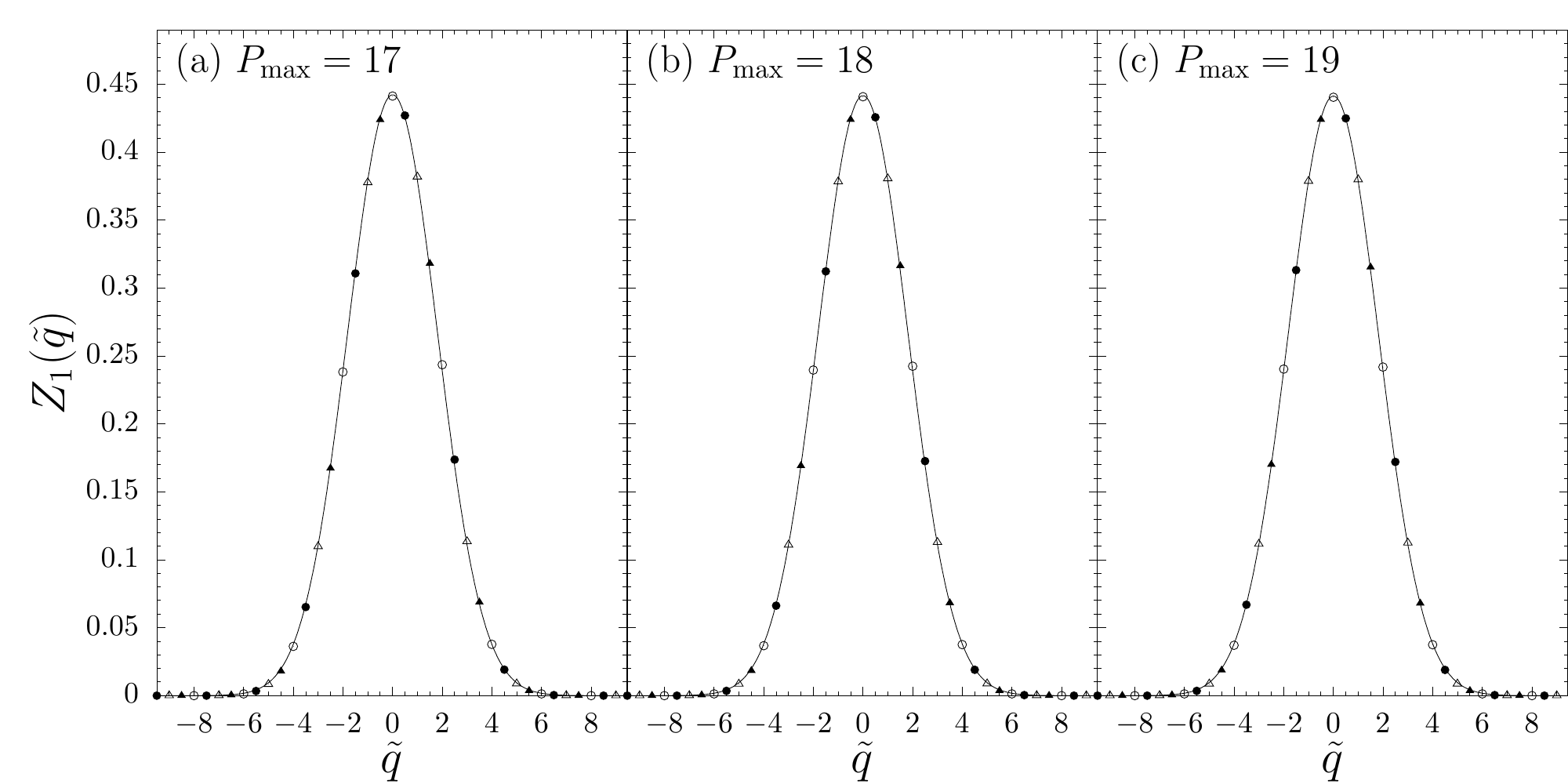}
\caption{FCS for the Laughlin $\nu=1/2$ state on a cylinder with perimeter $L=23$ (this is slightly above the range shown in Fig.~\ref{fig:laughlinhalfchargefluctuations} for pedagogical purposes) for CFT truncation parameters (a) $P_{\rm max}=17$, (b) $P_{\rm max}=18$ and (c) $P_{\rm max}=19$. The symbols are identical to those of Fig.~\ref{fig:laughlinhalfchargefluctuations}. As can be observed, the asymmetry around $\tilde{q}=0$ is more pronounced at the lowest $P_{\rm max}$ but is still visible at $P_{\rm max}=19$.}\label{app:fig:chargefluctuationvspmax}
\end{figure}

\begin{figure}[!h]
\centering
\includegraphics[width=0.99\columnwidth]{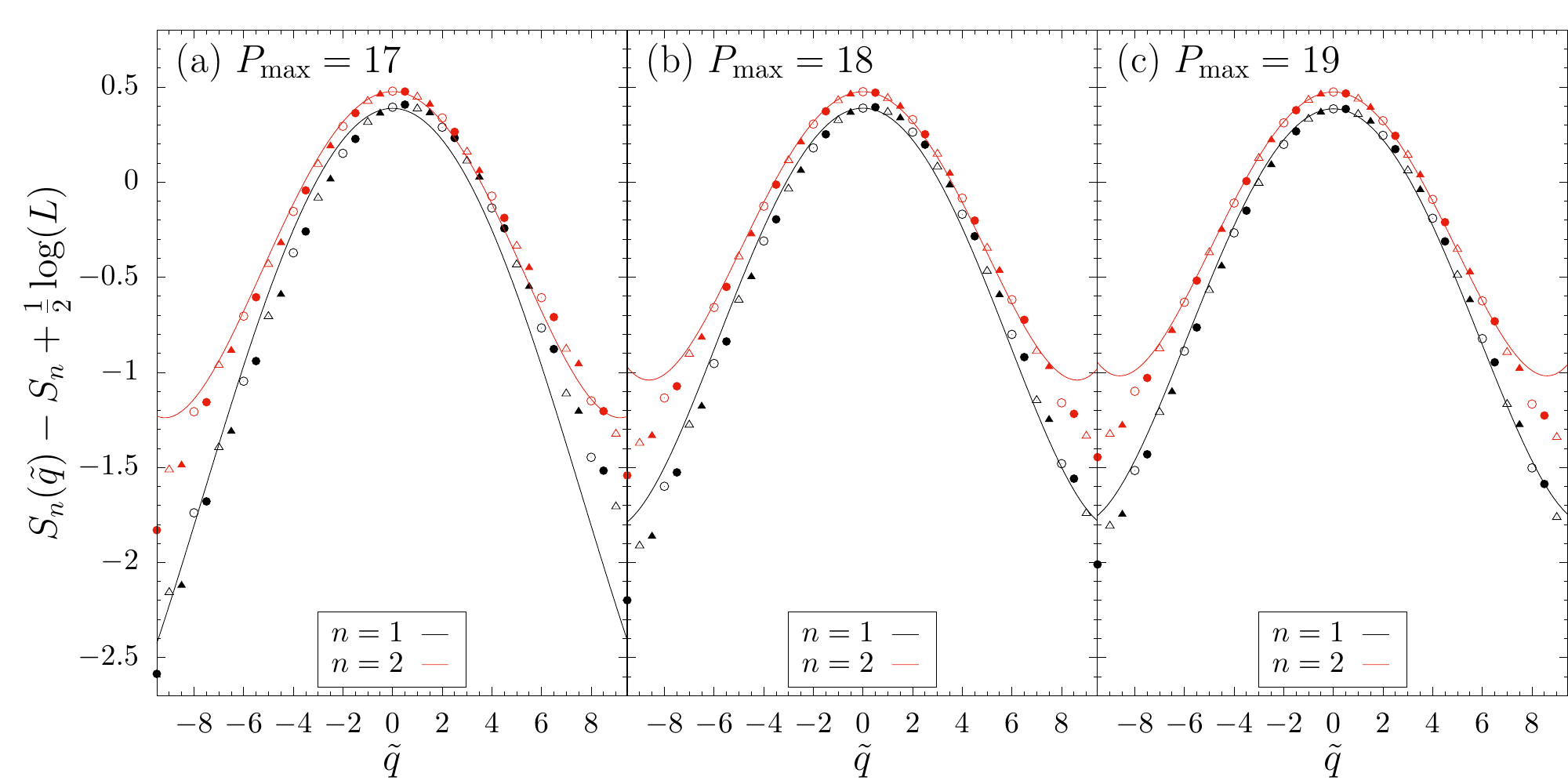}
\caption{Charge dependence of EE for the Laughlin $\nu=1/2$ state on a cylinder with perimeter $L=23$ (again slightly larger than those of Fig.~\ref{fig:laughlinhalfchargefluctuations} for pedagogical purposes) for CFT truncation parameters (a) $P_{\rm max}=17$, (b) $P_{\rm max}=18$ and (c) $P_{\rm max}=19$. For each truncation, we provide both the von Neumann entropy ($n=1$ in black) and the second R\'enyi entropy ($n=2$ in red).  The symbols are identical to those defined in Fig.~\ref{fig:laughlinhalfchargefluctuations}. A strong asymmetry around $\tilde{q}=0$ is observed at the lowest $P_{\rm max}$ and step-like effects plague large $|\tilde{q}|$ values, both effects disappearing when $P_{\rm max}$ increases.}\label{app:fig:entropyvspmax}
\end{figure}

As mentioned in the main text, the finite truncation in the MPS leads to deviations in both FCS and charge-resolved EE. In Fig.~\ref{app:fig:chargefluctuationvspmax}, we show the  FCS at fixed perimeter for three different values of $P_{\rm max}$. Asymmetries around $\tilde{q}=0$ occur when this truncation is not large enough. Note that the charge fluctuation is a dominant contribution as opposed to the $\tilde{q}^2/L$ corrections to the charge-resolved EE considered \eg in Fig.~\ref{fig:laughlinhalfentropy}. As such, the former is more robust than the latter. Figure \ref{app:fig:entropyvspmax} exemplifies this difference. It shows the $\tilde{q}^2/L$ corrections to the charge-resolved EE at fixed $L$ and different $P_{\rm max}$. The insufficient truncation manifests itself as an asymmetry around $\tilde{q}=0$ and step-like effect at larger $\tilde{q}$, and is more prominent for von Neumann EE than the second R\'enyi entropy.

\bibliographystyle{apsrev4-1}
\bibliography{biblio}

\end{document}